\documentclass[12pt]{article}

\usepackage{amsmath}
\usepackage{amsfonts}
\usepackage{amssymb}
\usepackage{amsthm}
\usepackage{graphicx}
\usepackage{xcolor}
\usepackage{float}
\usepackage{booktabs}
\usepackage{multirow}
\usepackage{url}
\usepackage{nicefrac}
\usepackage[expansion=false]{microtype}
\usepackage{caption}
\usepackage{subcaption}
\usepackage{algorithm}
\usepackage{algpseudocode}
\usepackage{tikz}
\usetikzlibrary{arrows.meta, positioning, fit, backgrounds, calc, shapes, shapes.geometric, decorations.pathreplacing, calligraphy}
\usepackage{neuralnetwork}
\usepackage{stmaryrd}
\usepackage{bbold}
\usepackage{pgfplots}
\usepackage{pgfplotstable}
\usepackage{adjustbox}

\graphicspath{{./images/}{./figs/}{./cc_plots/}{./tucker_spline_plots/}{./hybrid_plots/}}

\newcommand{\totepochs}{19294}
\newcommand{\checkinterval}{500}

\newcommand{\warmupsteps}{100}
\newcommand{\decayinterval}{2000}

\newcommand{\clipnorm}{1.0}
\newcommand{\initgain}{0.1}

\newcommand{\numlayers}{16}

\begin{document}

\title{Hybrid Neural Interpolation of a Sequence of Wind Flows}

\author{
Ameir Shaa$^{1,*}$, 
Claude Guet$^{1}$, \\ 
Xiasu Yang$^{2}$, 
Armand Albergel$^{2}$, 
Bruno Ribstein$^{2}$, 
Maxime Nibart$^{2,\dagger}$
}

\date{}

\maketitle

\vspace{-1em}
\begin{center}
\small $^{1}$School of Physical and Mathematical Sciences, Nanyang Technological University,\\
\small 21 Nanyang Link, Singapore 637371\\[0.5em]
\small $^{2}$SUEZ ARIA Technologies, 15 Rue du Port, 92000 Nanterre, France\\[1em]
\small $^{*}$Corresponding author: \texttt{ameirshaa.akberali@ntu.edu.sg}\\
\small Formerly (Sep 2023 to Nov 2024) at CNRS@CREATE Ltd., 1 CREATE Way, \#08-01,\\
\small CREATE Tower, Singapore 138602\\[0.5em]
\small $^{\dagger}$Now deceased\\[0.5em]
\small Other emails: \texttt{cguet@ntu.edu.sg}, \texttt{xiasu.yang@suez.com},\\
\small \texttt{armand.albergel@suez.com}, \texttt{bruno.ribstein@suez.com}
\end{center}
\vspace{1em}

\begin{abstract}
Rapid and accurate urban wind field prediction is essential for modeling particle transport in emergency scenarios. Traditional Computational Fluid Dynamics (CFD) approaches are too slow for real-time applications, necessitating surrogate models. We develop a hybrid neural interpolation method for constructing surrogate models that can update urban wind maps on timescales aligned with meteorological variations.

Our approach combines Tucker tensor decomposition with neural networks to interpolate Reynolds-Averaged Navier-Stokes (RANS) solutions across varying inlet wind angles. The method decomposes high-dimensional velocity, pressure, and eddy viscosity field datasets into a core tensor and factor matrices, then uses Fourier interpolation for angular modes and k-nearest neighbors convolution for spatial interpolation. A neural network correction mitigates interpolation artifacts while preserving physical consistency.

We validate the approach on a simple cylinder-sphere configuration and, relative to a strong pure neural network benchmark, achieve comparable or improved accuracy ($R^2 > 0.99$) with significantly reduced training time. The pure NN remains a feasible reference model; the hybrid provides an accelerated approximate alternative that suppresses spurious oscillations, maintains wake dynamics, and demonstrates computational efficiency suitable for real-time urban wind simulation.
\end{abstract}

\noindent\textbf{Keywords:} Urban wind modeling, Computational fluid dynamics, Neural networks, Tucker decomposition, RANS equations, Surrogate modeling, Wind field interpolation

\section{Introduction}

Accurately predicting urban wind flows remains a formidable challenge due to the geometric complexity of buildings, streets, and vegetation, which generate intricate flow separation and vortex interactions \cite{britter2003flow,franke2007best}. These wake interactions lead to strongly nonlinear flow dynamics. Yet, accurate prediction is essential for applications such as pollutant dispersion monitoring and rapid inverse identification of emission sources, where fast, reliable flow field estimates are critical.

The incompressible Navier–Stokes equations \cite{batchelor1967,temam2001} fully describe the underlying physics, but Direct Numerical Simulation (DNS) \cite{moin1998} is computationally infeasible for realistic urban domains. A common simplification is to apply time-averaging on meteorological timescales ($\sim 15$ min) to obtain the Reynolds-Averaged Navier–Stokes (RANS) equations \cite{Pope2000,reynolds1895}, supplemented by turbulence closures. The steady-state RANS solution provides averaged velocity, eddy viscosity and pressure fields that capture dominant transport pathways. Nevertheless, RANS computations remain costly, especially for high-resolution unstructured meshes or large domains.

This work seeks to develop a surrogate model capable of reproducing the RANS wind and eddy viscosity fields within minutes on a standard laptop, given a set of boundary parameters. Here, we focus on the inlet wind angle as the varying parameter, while maintaining a fixed vertical profile. High-fidelity training data are generated from Code\_Saturne simulations on a canonical cylinder–sphere geometry. The main challenge lies in interpolating the three-dimensional flow fields across inlet angles, where nonlinear transitions in wake topology occur—for instance, from shielded to exposed obstacle configurations. Simple linear interpolation or regression fails to reproduce these nonlinear transitions faithfully.

Classical reduced-order modeling via Proper Orthogonal Decomposition (POD) \cite{berkooz1993proper,du2013pod} achieves efficiency by projecting CFD snapshots onto a low-dimensional subspace of dominant modes. However, because POD constrains solutions to a fixed linear subspace, it performs poorly when boundary variations induce distinct flow regimes, such as those arising from changes in wind angle \cite{amsallem2008interpolation}. The resulting nonlinear deformations of the solution manifold cannot be captured within a single basis, motivating the need for parametric or nonlinear surrogates.

Recent neural network (NN) approaches offer greater flexibility. Physics-Informed Neural Networks (PINNs) \cite{raissi2019physics,karniadakis2021,cuomo2022} incorporate governing equations as soft constraints, while purely data-driven NNs \cite{lee2020data} directly learn input–output mappings from CFD databases. However, PINNs face challenges at high Reynolds numbers and require delicate loss balancing, whereas data-driven models, though achieving high accuracy, entail significant training costs ($\approx 2.82$ s per epoch in our experiments) and large parameter counts (50 k even for simple geometries). While feasible for establishing accuracy benchmarks, these computational demands motivate the development of more efficient alternatives. In this work, the pure NN serves as the accuracy benchmark (Section~\ref{sec:purenn}), while the hybrid provides an accelerated approximate alternative (Section~\ref{sec:hybridnn}).

Tensor decomposition methods, such as Tucker decomposition \cite{Kolda2009,hitchcock1927,carroll1970}, offer a structured means of reducing dimensionality while retaining multi-modal interactions. Tucker decomposition factorizes a high-dimensional field into a compact core tensor and orthonormal factor matrices, enabling efficient storage and parameter interpolation. Yet, when applied to urban wind flows, standard interpolation across angular parameters often produces Gibbs-like oscillations due to discontinuities in the solution manifold \cite{amsallem2008interpolation,reiss2018shifted,peherstorfer2018survey,kutz2016dmd}.

To address these limitations, we develop a hybrid tensor–neural framework that embeds a Tucker-decomposed ansatz directly within a neural network architecture. The tensor representation provides structured dimensionality reduction (compressing 100,680 grid points into 120 spatial modes), while the neural network learns residual corrections that suppress interpolation artifacts and restore physical consistency. Unlike previous tensor–NN hybrids that use decomposition solely as preprocessing, our approach integrates the tensor structure into the forward pass via a bottleneck layer and augments it with a self-consistent residual learning module.

Our paper makes three primary contributions: (1) We develop a hybrid surrogate model that combines Tucker tensor decomposition with neural network correction, specifically addressing interpolation artifacts that plague pure POD methods; (2) We introduce a self-consistent residual learning framework with structured perturbations that suppresses spurious oscillations while maintaining physical consistency; (3) We demonstrate significant training acceleration through architectural efficiency—reducing parameter count from $\sim$50k to $\sim$16k—while achieving superior accuracy ($R^2 > 0.99$) on complex wake interactions. Unlike existing tensor–NN combinations that use tensor decompositions primarily for preprocessing or network compression \cite{Kolda2009,novikov2015tensorizing,kossaifi2019tensor}, our method embeds the tensor ansatz directly into the network architecture, allowing the latent tensor structure to participate in the forward dynamics.

The remainder of this paper is organized as follows. Section 2 describes the RANS dataset generation and highlights the limitations of direct CFD-based modeling for urban domains. Section 3 presents the surrogate modeling framework, including the direct NN baseline and the proposed hybrid Tucker-NN model as an accelerated approximate alternative. Section 4 provides the combined Discussion and Conclusion.

For clarity, we use lowercase letters for scalars ($a,b,\ldots$), 
uppercase for vectors ($U, V, \ldots$), 
bold uppercase for matrices ($\mathbf{A}, \mathbf{B}, \ldots$), 
and calligraphic letters for tensors ($\mathcal{A}, \mathcal{B}, \ldots$). 
Einstein summation convention applies unless otherwise stated.

\section{Production of RANS Wind Field Training Data}

Consider a wind velocity field, denoted by 
\[
U = [u_x, u_y , u_z ]^T,
\]
that obeys the Navier–Stokes equations
\begin{align}
\frac{\partial U}{\partial t} + (U \cdot \nabla) U &= -\frac{1}{\rho}\nabla p + \nu \nabla^2 U, \\
\nabla \cdot U &= 0, 
\end{align}
where the density $\rho$ is assumed to be constant, $p$ is the static pressure, and $\nu$ is the kinematic viscosity. To obtain the Reynolds-averaged Navier-Stokes (RANS) equations \cite{Pope2000,reynolds1895} we decompose the time-dependent velocity into a mean time-averaged component denoted for simplicity by $U$ and a fluctuating component $U'$ which averages to zero:
\begin{align}
 u_i(R, t) = u_i(R) + u'_i(R, t), \\
p(R, t) = p(R) + p'(R, t).
\end{align}
where $R = [x, y, z]^T$ and $(R)_i = r_i$. 

After time-averaging all terms, the steady RANS equations read as:
\begin{align}
u_j\frac{\partial u_i}{\partial r_j} = -\frac{1}{\rho}\frac{\partial p}{\partial r_i} + \nu \frac{\partial^2 u_i}{\partial r_j \partial r_j} - \frac{\partial}{\partial r_j}\overline{u_i' u_j'} ,
\end{align}
\begin{align}
\frac{\partial u_i}{\partial r_i} = 0.
\end{align}
where the Reynolds stress tensor, $\tau_{ij} = \overline{u'_i u'_j}$, represents the additional stresses due to turbulent fluctuations. In practice, these stresses are described through the $(k\text{-}\varepsilon)$ turbulence model \cite{jones1972,launder1974numerical,rodi1993} to close the system of equations to yield;
\begin{align}
\text{(Continuity)}
\quad & \frac{\partial u_i}{\partial r_i} = 0 \\[1.2ex]
\text{(Momentum)}
\quad & u_j \frac{\partial u_i}{\partial r_j}
- \frac{\partial}{\partial r_j} \left[ (\nu + \nu_t) \left( \frac{\partial u_i}{\partial r_j} + \frac{\partial u_j}{\partial r_i} \right) \right]
= -\frac{\partial P^*}{\partial r_i} \\[1.2ex]
\text{(TKE: $k$)}
\quad & u_j \frac{\partial k}{\partial r_j}
- \frac{\partial}{\partial r_j} \left( \frac{\nu_t}{\sigma_k} \frac{\partial k}{\partial r_j} \right)
= \nu_t \left( \frac{\partial u_i}{\partial r_j} + \frac{\partial u_j}{\partial r_i} \right) \frac{\partial u_i}{\partial r_j} - \varepsilon \\[1.2ex]
\text{(Dissipation: $\varepsilon$)}
\quad & u_j \frac{\partial \varepsilon}{\partial r_j}
- \frac{\partial}{\partial r_j} \left( \frac{\nu_t}{\sigma_\varepsilon} \frac{\partial \varepsilon}{\partial r_j} \right)
= c_1 \nu_t \frac{\varepsilon}{k} \left( \frac{\partial u_i}{\partial r_j} + \frac{\partial u_j}{\partial r_i} \right) \frac{\partial u_i}{\partial r_j}
- c_2 \frac{\varepsilon^2}{k} \\[1.2ex]
\text{(Eddy viscosity)}
\quad & \nu_t = c_\mu \frac{k^2}{\varepsilon} \\[1.2ex]
\text{(Modified pressure)}
\quad & P^* = P + \frac{2}{3}k
\end{align}

\begin{equation}
 u_j \frac{\partial u_i}{\partial r_j} =
- \frac{1}{\rho} \frac{\partial p}{\partial r_i} - \frac{1}{\rho} \frac{\partial}{\partial r_i} \left( \frac{2}{3} \rho k \right) +
\frac{\partial}{\partial r_j} \left[ \left( \nu + C_\mu \frac{k^2}{\varepsilon} \right) \frac{\partial u_i}{\partial r_j} \right]
\end{equation}
where $k$ is the turbulent kinetic energy (TKE), $\varepsilon$ is the turbulent kinetic energy dissipation rate, $\nu_t$ is the eddy viscosity. 

In this study, the open source Finite-Volume CFD software Code\_Saturne~\cite{archambeau2004codesaturne} is used to solve the RANS equations with the standard $k-\varepsilon$ turbulence to produce training datasets of wind fields. The finite volume method \cite{ferziger2002} provides spatial discretization. The constants in the $k-\varepsilon$ model given by \cite{launder1974numerical} are listed in Table~\ref{tab:k-e_cst}.

\begin{table}[htbp]
	\caption{Constants in the standard $k-\varepsilon$ turbulence model}
	\centering
	\begin{tabular}{lllll}
		\toprule
		$c_{\mu}$ & $c_1$ & $c_2$ & $\sigma_k$ & $\sigma_{\varepsilon}$ \\
		\midrule
		0.09 & 1.44 & 1.92 & 1.0 & 1.3 \\
		\bottomrule
	\end{tabular}
	\label{tab:k-e_cst}
\end{table}

\subsection{Test Model Geometry} \label{subsec:cylinder}

\begin{figure}[htbp]
  \centering
  \begin{subfigure}[b]{0.3\textwidth}
    \centering
    \includegraphics[width=\textwidth]{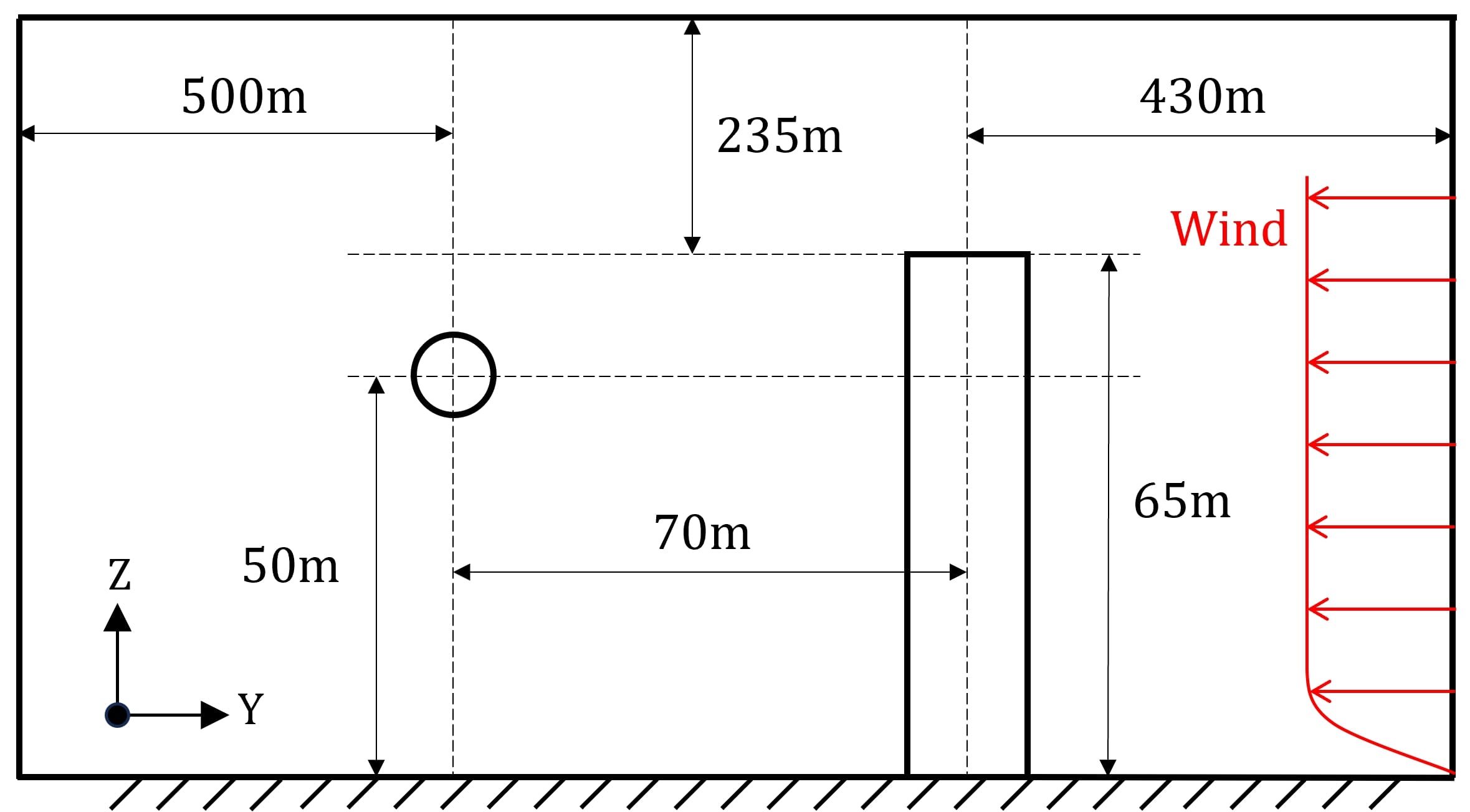}
    \caption{}
    \label{subfig:cylinder_sphere_side}
  \end{subfigure}
  \hfill
  \begin{subfigure}[b]{0.3\textwidth}
    \centering
    \includegraphics[width=\textwidth]{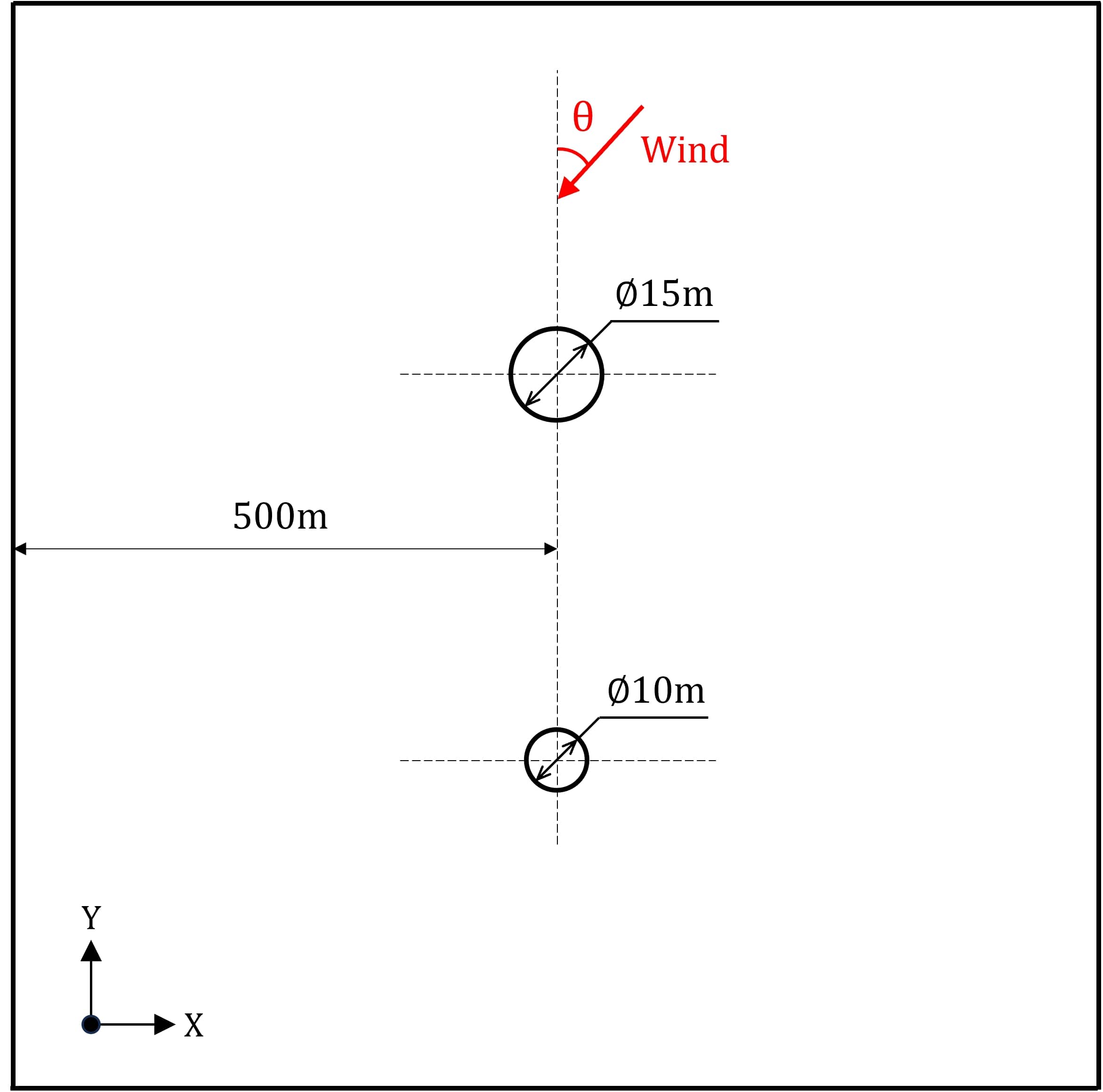}
    \caption{}
    \label{subfig:cylinder_sphere_top}
  \end{subfigure}
  \hfill
  \begin{subfigure}[b]{0.3\textwidth}
    \centering
    \includegraphics[width=\textwidth]{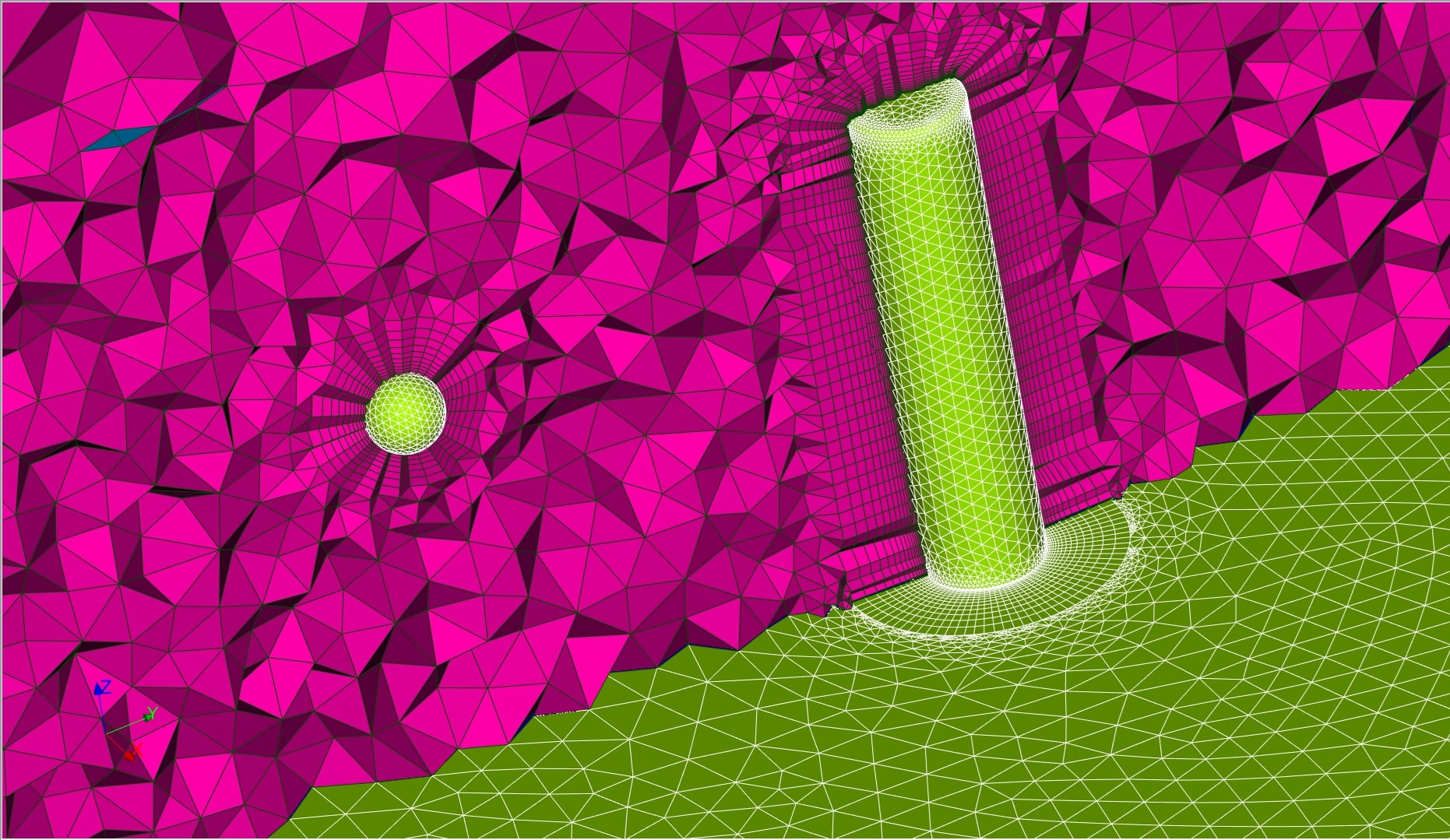}
    \caption{}
    \label{subfig:cylinder_sphere_mesh}
  \end{subfigure}
  \caption{Cylinder-sphere test configuration: 
  (a) Side view showing a wall-mounted cylinder (15 m diameter, 65 m height) and a floating sphere (10 m diameter, centered at 50 m elevation); 
  (b) Top view of the configuration showing the relative positioning of obstacles. The inlet wind angle $\theta$ is measured counterclockwise from the positive x-axis; 
  (c) Computational grid for the cylinder-sphere test case, consisting of 317,551 cells. The domain is discretized using tetrahedral elements with prismatic inflation layers near solid surfaces to resolve boundary layer flows. Mesh refinement is concentrated in the wake regions.}
  \label{fig:cylinder_sphere_top_side}
\end{figure}

In this section, we present an academic test-case consisting of a regular geometrical configuration where a sphere with a diameter of 10 m floating at 50 m above ground is positioned next to a wall-mounted cylinder with a diameter of 15 m and a height of 65 m. The cylinder is centered at $(x, y) = (500, 570)$ m and the sphere at $(x, y) = (500, 500)$ m, resulting in a center-to-center horizontal separation of 70 m. The side and top views of this test configuration are illustrated in Fig.~\ref{subfig:cylinder_sphere_side} and Fig.~\ref{subfig:cylinder_sphere_top}.

The computational grid shown in Fig.~\ref{subfig:cylinder_sphere_mesh} is generated with the open source CAD software SALOME~\cite{salome}. The bulk of the 3D domain is discretized using tetrahedral cells while prismatic inflation layers are created from the solid surfaces in order to correctly model the near-wall flows. A total of 317,551 cells are included in the computational grid. 

The wind speed, turbulence kinetic energy (TKE) and dissipation rate profiles for a neutral stability condition are constructed for the incoming wind by following the recommended procedure described in \cite{lacome2017guide}. These profiles are based on atmospheric boundary layer theory \cite{panofsky1984,stull1988}. The reference wind speed at 10 m above ground is set to 2 m/s, the Monin-Obukhov length scale is set to $10^4$ m with a roughness length of 1 m to simulate the condition in urban areas. 


At the height of the sphere's centroid ($z = 50$ m) the wind speed $u = 3.4$ m/s which translates to a diameter-based Reynolds number $Re_d = 2.3 \times 10^6$ for the sphere and $Re_d = 3.5 \times 10^6$ for the cylinder. At these high Reynolds numbers, the wakes behind the two obstacles become narrow and highly turbulent with no time-dependent periodic vortex streets \cite{achenbach1974vortex,leishman2023bluff,roshko1961experiments,williamson1996}, which makes the steady-state RANS modeling a suitable approach for the current CFD simulations.


By rotating the incoming wind profile from $0^{\circ}$ to $345^{\circ}$ with a step of $15^{\circ}$, a total of 24 RANS simulations are performed to generate wind fields as a function of the inlet wind angle $\theta$. Two additional test cases at intermediate angles $\theta= 7.5^{\circ}$ and $157.5^{\circ}$ are not included in the training dataset and serve as held-out test angles for model evaluation.

\begin{figure}[h!]
  \centering
  \begin{subfigure}[b]{0.23\textwidth}
    \centering
    \includegraphics[width=\textwidth]{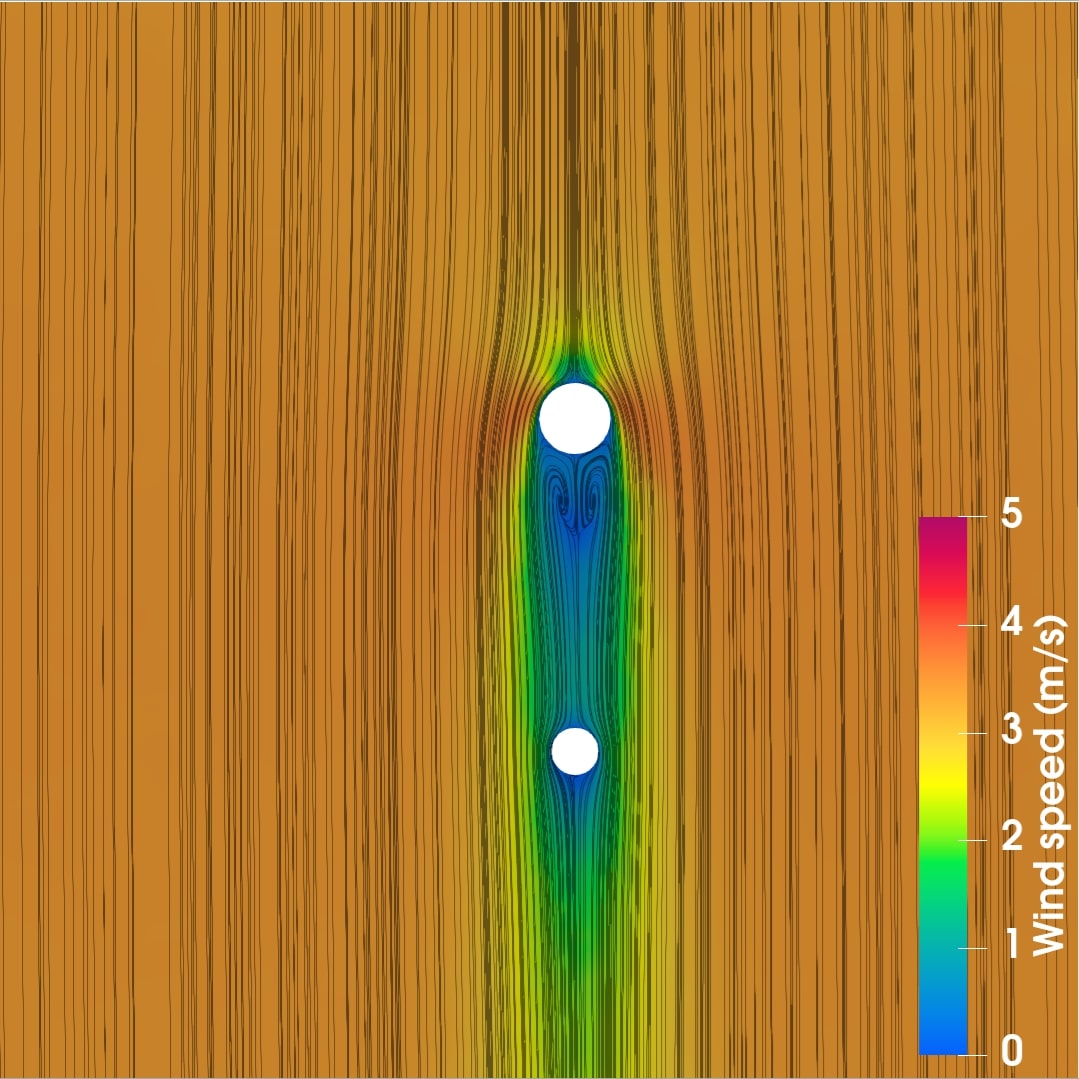}
    \caption{$\theta=0^\circ$}
    \label{subfig:cylinder_sphere_zslices_0}
  \end{subfigure}
  \hfill
  \begin{subfigure}[b]{0.23\textwidth}
    \centering
    \includegraphics[width=\textwidth]{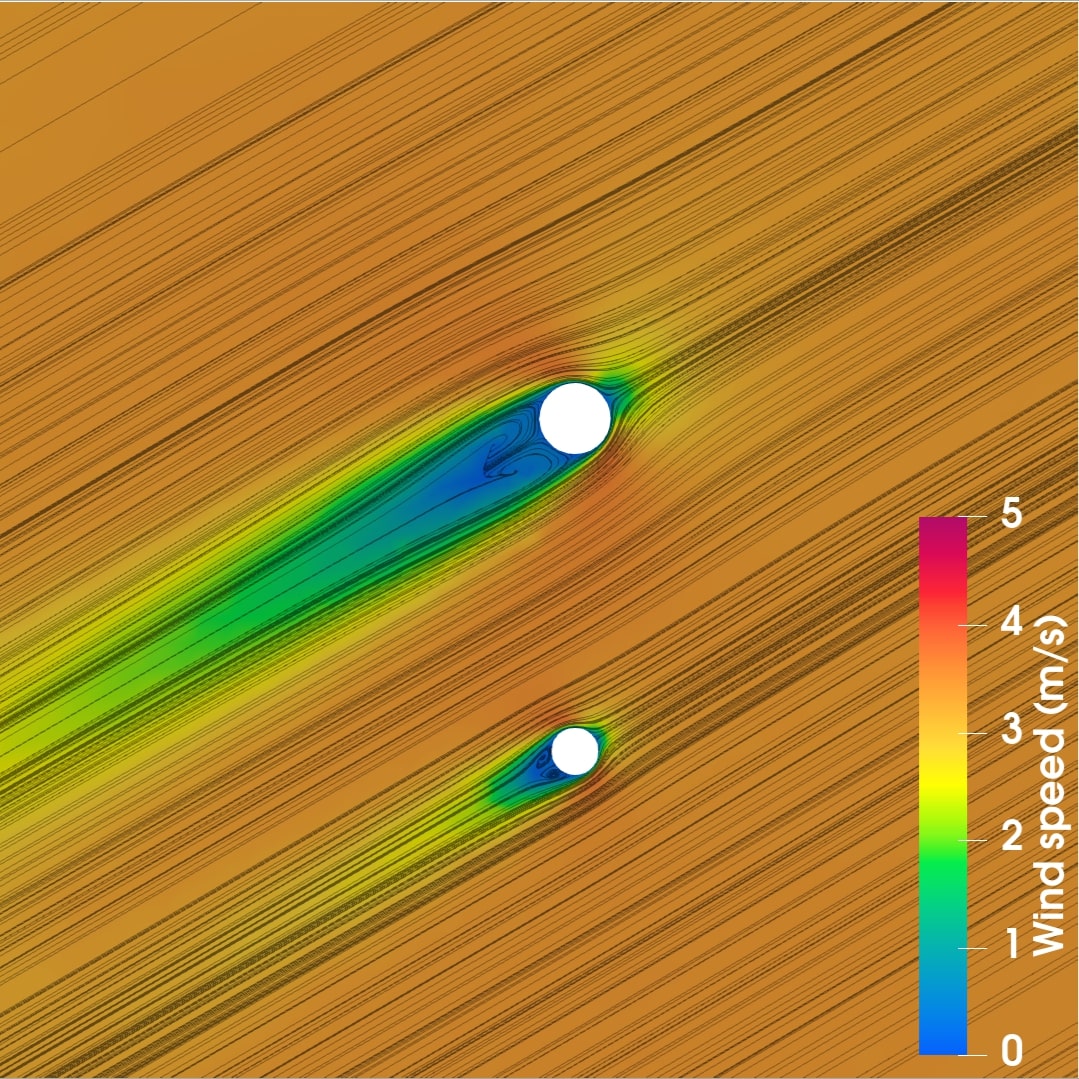}
    \caption{$\theta=60^\circ$}
    \label{subfig:cylinder_sphere_zslices_60}
  \end{subfigure}
  \hfill
  \begin{subfigure}[b]{0.23\textwidth}
    \centering
    \includegraphics[width=\textwidth]{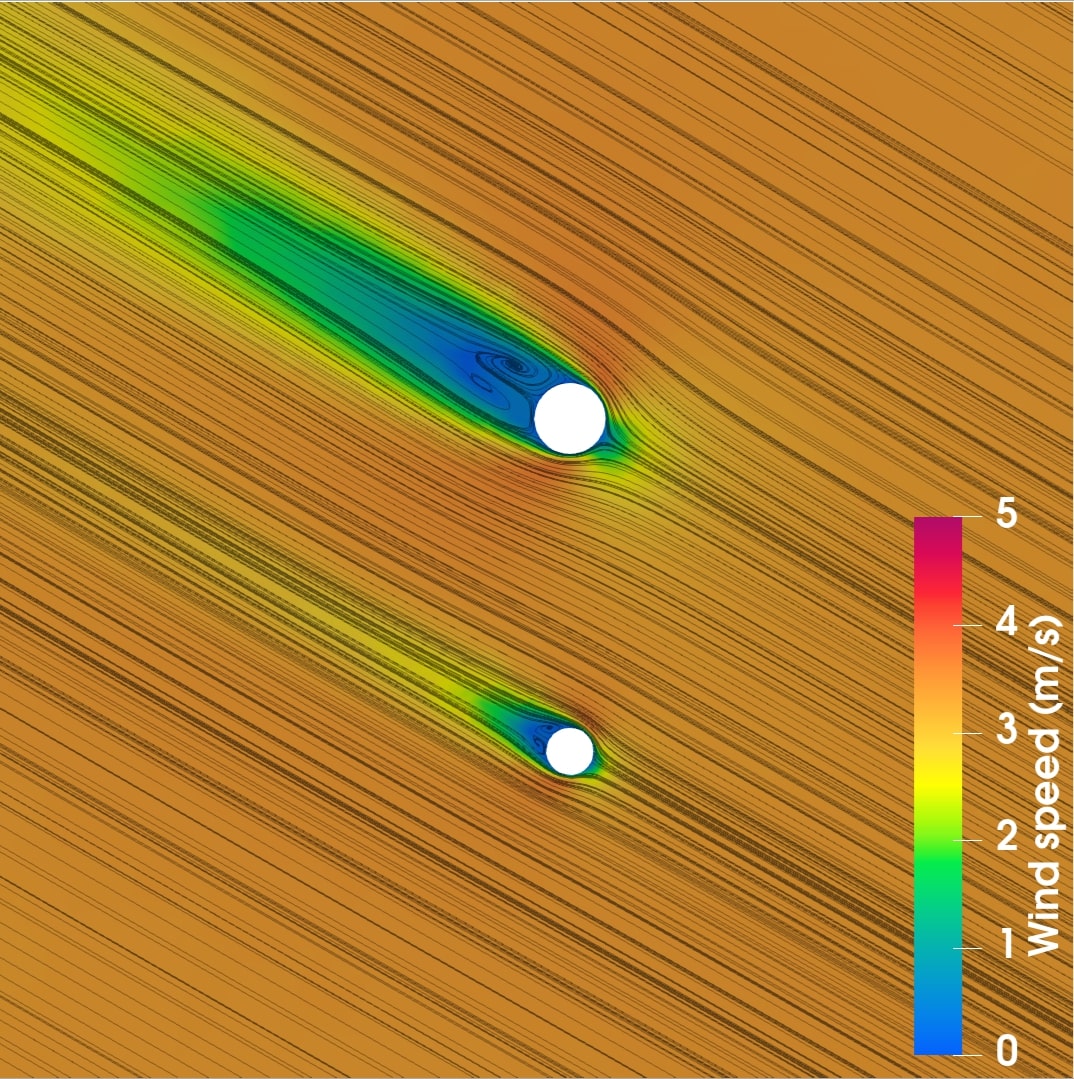}
    \caption{$\theta=120^\circ$}
    \label{subfig:cylinder_sphere_zslices_120}
  \end{subfigure}
  \hfill
  \begin{subfigure}[b]{0.23\textwidth}
    \centering
    \includegraphics[width=\textwidth]{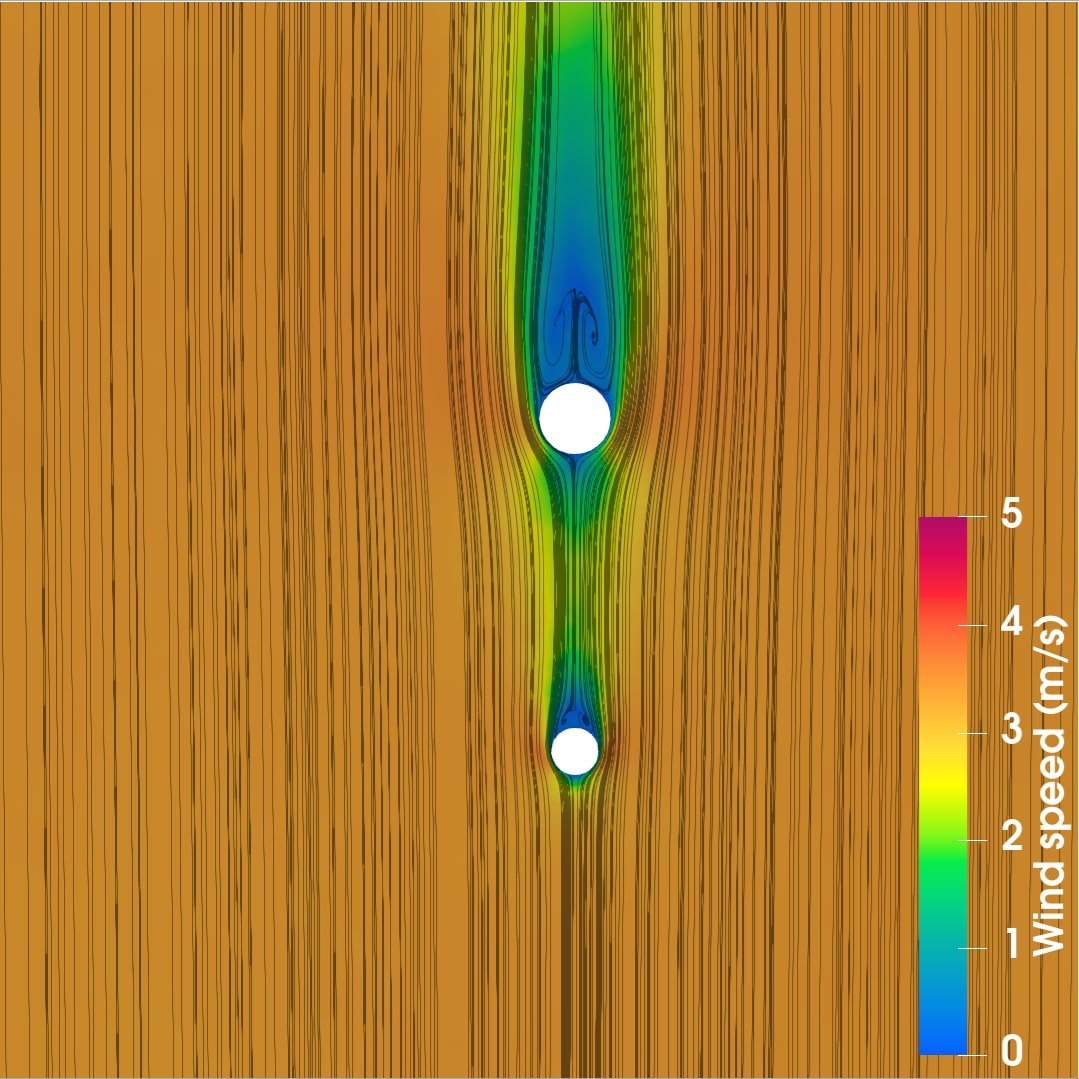}
    \caption{$\theta=180^\circ$}
    \label{subfig:cylinder_sphere_zslices_180}
  \end{subfigure}

  \caption{Wind speed magnitude and streamlines on the horizontal plane at $z=50\,\mathrm{m}$ for the cylinder-sphere configuration. 
  (a) $\theta=0^\circ$: sphere shielded by cylinder, wake interaction suppresses sphere vortices; 
  (b) $\theta=60^\circ$: independent wake development with dual vortex pairs; 
  (c) $\theta=120^\circ$: similar to $\theta=60^\circ$ but rotated; 
  (d) $\theta=180^\circ$: cylinder minimally affected by sphere in the upstream due to size difference. Streamlines are all in black, they are however overlaid on a horizontal slice colored by velocity magnitude.}
  \label{fig:cylinder_sphere_zslices}
\end{figure}

In Fig.~\ref{fig:cylinder_sphere_zslices}, we can visualize the wind speed maps and surface streamlines for $\theta\in \{0^\circ, 60^\circ, 120^\circ, 180^\circ\}$ on a 2D slice cutting through the horizontal symmetry plane of the sphere at the height of 50 m.

Although the current test case has a relatively simple geometry, some non-linear flow patterns can still be observed with varying inlet wind angle due to the interference between the two wakes formed behind the cylinder and the sphere. When the two wake regions remain relatively unperturbed by each other ($\theta\in \{60^\circ, 120^\circ\}$), they tend to develop highly similar flow patterns that extend to downwind distances proportional to the respective diameters of their generating obstacles. For these two inlet angles, we observe two counter rotating vortices on the leeward sides of both objects due to vortex shedding. However, this classical vortical structure is disrupted at $\theta=0^\circ$ when the smaller sphere is shielded by the larger cylinder along the flow path as shown in Fig.~\ref{subfig:cylinder_sphere_zslices_0}. The vortices behind the sphere disappear under the influence of the wake region generated by the cylinder upstream. On the other hand, when we reverse the incoming wind at $\theta=180^\circ$ so that the cylinder is shielded by the sphere (Fig.~\ref{subfig:cylinder_sphere_zslices_180}), the disappearance of the two counter-rotating vortices behind the cylinder does not take place as one might expect given the streamline pattern at $\theta=0^\circ$.

\begin{figure}[h!]
  \centering
  \begin{subfigure}[b]{0.45\textwidth}
    \centering
    \includegraphics[width=\linewidth]{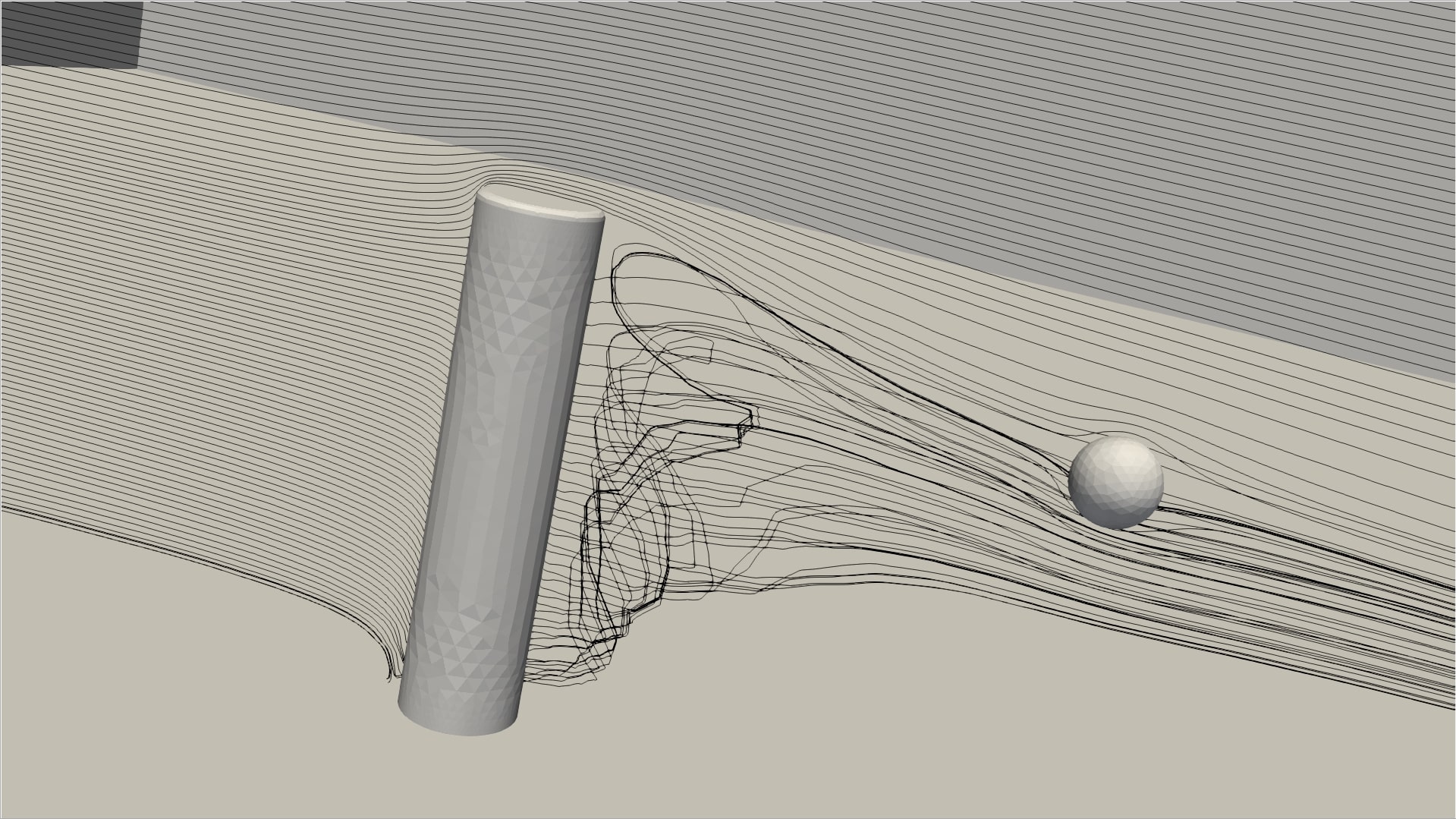}
    \caption{}
    \label{subfig:cylinder_sphere_stream_0deg}
  \end{subfigure}
  \hfill
  \begin{subfigure}[b]{0.45\textwidth}
    \centering
    \includegraphics[width=\linewidth]{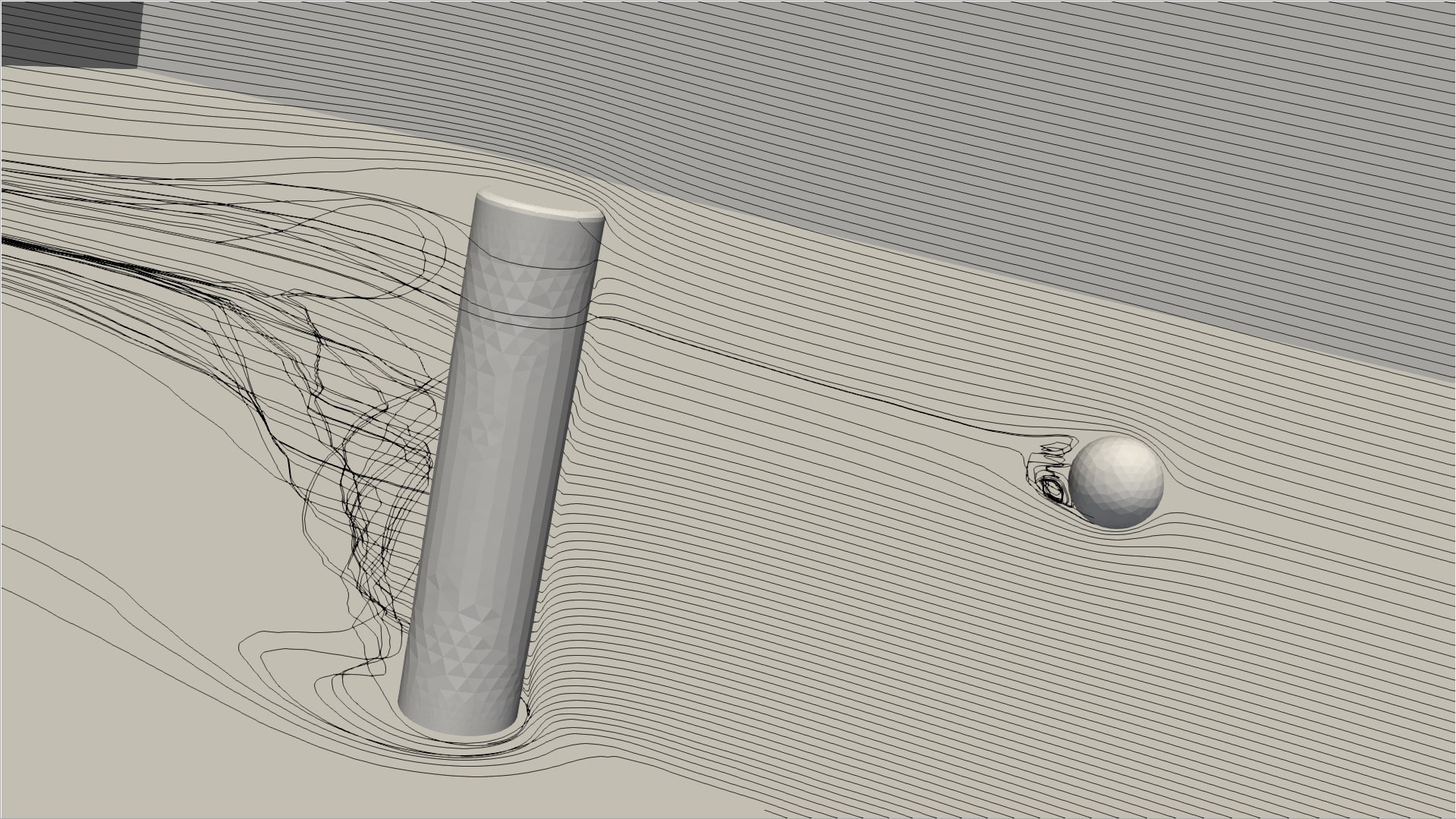}
    \caption{}
    \label{subfig:cylinder_sphere_stream_180deg}
  \end{subfigure}

  \caption{Three-dimensional streamlines for the cylinder-sphere configuration:  
  (a) $\theta = 0^\circ$ — sphere positioned downstream of the cylinder. The sphere is completely enclosed within the cylinder’s turbulent wake, suppressing the formation of independent sphere vortices;  
  (b) $\theta = 180^\circ$ — cylinder positioned downstream of the sphere. The smaller sphere wake has minimal influence on the cylinder’s wake structure due to the size difference.}
  \label{fig:cylinder_sphere_streamlines}
\end{figure}

The wake interactions can be more clearly illustrated with 3D streamlines in Fig.~\ref{subfig:cylinder_sphere_stream_0deg} and Fig.~\ref{subfig:cylinder_sphere_stream_180deg}. The size difference between the wakes behind the cylinder and the floating sphere becomes apparent by observing the 3D vortical structures. At $\theta=0^\circ$ (Fig.~\ref{subfig:cylinder_sphere_stream_0deg}), the sphere is completely enclosed in the turbulent wake region behind the cylinder, whereas, at $\theta=180^\circ$ (Fig.~\ref{subfig:cylinder_sphere_stream_180deg}), we observe that the wake generated by the sphere causes such a limited impact on the downwind cylinder that the structure of its wake region remains practically unchanged compared with the one in Fig.~\ref{subfig:cylinder_sphere_stream_0deg}. This observation suggests potential generalization to real-world urban wind fields in the presence of buildings with varying dimensions: a change in the positioning of the buildings relative to the wind direction can trigger a fundamental and unpredictable transformation in the structures of the turbulent wakes.

\section{Surrogate Models}
Having in hand a high-fidelity model for the wind flow providing us with high-dimensional $(u_x, u_y, u_z, p, \nu_t)$ data on an unstructured grid at 24 inlet wind angles ranging equally from $0^{\circ}$ to $345^{\circ}$, we have developed two distinct surrogate models with complementary objectives.

The first is a purely data-driven neural network regression model that serves as a reference benchmark for achievable accuracy, while the second is a faster, approximate alternative that combines Tucker tensor decomposition with a self-consistent neural correction to achieve comparable accuracy with significantly reduced computational cost. Both models are trained using the same angular sampling: all 24 inlet angles from $0^{\circ}$ to $345^{\circ}$ in $15^{\circ}$ increments. For the pure NN model (Section~\ref{sec:purenn}), we train on cell-centered values. For the hybrid Tucker-NN model (Section~\ref{sec:hybridnn}), we work with point (nodal) values obtained by interpolating cell data to mesh vertices. In both cases, the intermediate angles $7.5^{\circ}$ and $157.5^{\circ}$ are reserved as unseen evaluation angles to assess interpolation capability.

\subsection{Data-driven Neural Network Regression}\label{sec:purenn}

From each CFD simulation carried on at a given inlet wind angle \(\{\theta_i\}_{i=1}^{J}\), ($J=24$ in this section) we extract the corresponding 3D velocity field, scalar pressure field, and scalar eddy viscosity field using the open-source data analysis software ParaView \cite{Ahrens2005}. The CFD solver Code\_Saturne produces cell-centered fields on the computational mesh of 317,551 cells. For the pure NN model, we work directly with these cell-centered values, resulting in $N = 352{,}071$ data points per field (including boundary and internal cells).

For the sake of consistency and to ensure efficient neural network (NN) learning with dimensionless quantities, the output data are then normalized to a Gaussian distribution so that each field distribution has zero mean and unit variance as per Eq.~\eqref{eq:normal} below.

\begin{equation}
\label{eq:normal}
  \hat{x}_i \;=\; \frac{x_i - \mu}{\sigma},
  \quad
  \mu = \frac{1}{N}\sum_{j=1}^N x_j,
  \quad
  \sigma = \sqrt{\frac{1}{N}\sum_{j=1}^N (x_j - \mu)^2}.
\end{equation}

Then, all 5 dimensionless variables collected at 24 inlet angles from $0^\circ$ to $345^\circ$ in $15^\circ$ increments form the training data. Two intermediate angles, $7.5^\circ$ and $157.5^\circ$, are not used for training and are reserved for evaluation. The objective is to learn the mapping from spatial coordinates and wind angle to flow field variables, enabling predictions at arbitrary angles \(\theta\) and positions \((x, y, z)\) within the domain. 
The high-fidelity CFD data inherently satisfy conservation laws and boundary conditions. This motivates the use of a multi-layer perceptron (MLP) to identify patterns in the training data and interpolate them to intermediate inlet angles. However, this approach does not impose explicit physics-based constraints.

\subsubsection{NN Architecture and Optimization}
In this approach, the output is the five RANS field components:
\[
(u_x(R, \theta), u_y(R, \theta), u_z(R, \theta), p(R,\theta), \nu_t(R,\theta)),
\]
defined at any Cartesian position inside the domain and any inlet angle. 

\begin{figure}[htbp]
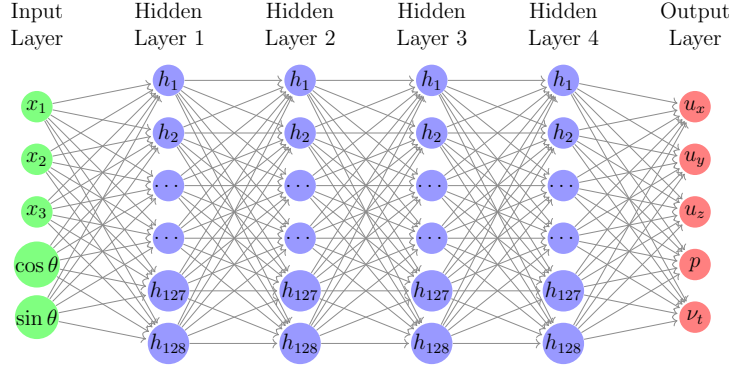

\centering
\scalebox{0.7}{
\begin{neuralnetwork}[height=6]

\newcommand{\x}[2]{\ifnum#2=4 $\cos\theta$\else\ifnum#2=5 $\sin\theta$\else $x_#2$\fi\fi}
\newcommand{\y}[2]{\ifcase#2\or $u_x$\or $u_y$\or $u_z$\or $p$\or $\nu_t$\fi}
\newcommand{\h}[2]{\ifnum#2=1 $h_1$\else\ifnum#2=6 $h_{128}$\else\ifnum#2=2 $h_2$\else\ifnum#2=5 $h_{127}$\else \dots\fi\fi\fi\fi}

\inputlayer[count=5, bias=false, title={Input Layer}, text=\x]
\hiddenlayer[count=6, bias=false, title={Hidden Layer 1}, text=\h] \linklayers
\hiddenlayer[count=6, bias=false, title={Hidden Layer 2}, text=\h] \linklayers
\hiddenlayer[count=6, bias=false, title={Hidden Layer 3}, text=\h] \linklayers
\hiddenlayer[count=6, bias=false, title={Hidden Layer 4}, text=\h] \linklayers
\outputlayer[count=5, title={Output Layer}, text=\y] \linklayers 

\end{neuralnetwork}
}
\caption{Architecture of the pure neural network regression model with four hidden layers of 128 neurons each, using ELU activation functions. Input layer receives 5 features: spatial coordinates ($x, y, z$) and encoded wind angle ($\cos\theta, \sin\theta$). Output layer predicts 5 RANS field variables: velocity components ($u_x, u_y, u_z$), pressure ($p$), and eddy viscosity ($\nu_t$). Total trainable parameters: $\sim$50,000.}
\label{fig:NN}
\end{figure}

The MLP architecture is shown in Fig.~\ref{fig:NN}. It comprises an input layer with 5 nodes $(x, y, z, \cos(\theta), \sin(\theta))$, four hidden layers each with 128 neurons, an output layer with 5 nodes corresponding to the five fluid components, and the Exponential Linear Unit (ELU) activation function \cite{Clevert2016} defined as follows. We intentionally use ELU in this \emph{pure NN baseline}, while the \emph{hybrid} model in Section~\ref{sec:hybridnn} uses ReLU within its residual/bottleneck path; the two activations are kept distinct by design to match each approach's objectives.

\begin{equation}\label{eq:ELU}
  f_\text{ELU}(x) = \begin{cases} 
  x & \text{if } x > 0 \\
  \alpha (e^x - 1) & \text{if } x \leq 0 
  \end{cases}
\end{equation}

with $\alpha=1$ being the default value.

The structure of the NN can be succinctly represented as:
\begin{equation}\label{eq:nnstructure}
  \begin{aligned}
    &\quad Z^{(l)} = \mathbf{W}^{(l)}A^{(l-1)} + B^{(l)}, \\
    &\quad A^{(l)} = f_\text{ELU}(Z^{(l)}), \quad \forall l \in \{1,2,3,4,5\}
  \end{aligned}
\end{equation}

where $\mathbf{W}^{(l)}$ is the weight matrix of layer $l$ with dimensions $n_l \times n_{l-1}$, and $B^{(l)}$ is the bias vector of size $n_l$.
The activation at layer $l$, denoted as $A^{(l)}$, is a vector of size $n_l$, with $A^{(0)}$ representing the input vector.
The pre-activation input to the neurons in layer $l$ is given by $Z^{(l)}$, which is transformed using the element-wise activation function $f_{\text{ELU}}(\cdot)$. The number of neurons per layer is $n_l = 128$ for $l = 1,2,3,4$ and $n_l = 5$ for $l = 0,5$. 

The neural network is trained using a dataset sparsely sampled from the high-fidelity RANS model. We adopt a training regime that includes dividing the dataset into training, validation, and testing subsets, employing a loss function that encapsulates only the fidelity to the CFD data (data loss). The data have been renormalized using the Standard Normal distribution. Training proceeds via backpropagation \cite{Rumelhart1986,lecun2012}.

The NN is built to predict the fluid velocity $\tilde{U}(R, \theta ; \mathbf{W}, B)$, the pressure $\tilde{p}(R, \theta; \mathbf{W}, B)$, and the eddy viscosity $\tilde{\nu_t}(R, \theta ; \mathbf{W}, B)$, where the symbol $\tilde{.}$ represents a function learned by a neural network. The loss function is defined as:

\begin{equation}
  \begin{aligned}
    \mathcal{L}_{\text{fit}} = \frac{1}{N} \sum_{i=1}^{N} \Big( & \|\tilde{U}(R_i, \theta_i ; \mathbf{W}, B) - U_i \|^2 \\
    & + \|\tilde{p}(R_i, \theta_i ; \mathbf{W}, B) - p_i \|^2 \\
    & + \|\tilde{\nu_t}(R_i, \theta_i ; \mathbf{W}, B) - \nu_{t_i}\|^2 \Big),
  \end{aligned}
\end{equation}
where $R_i$ is the cartesian position of the $i^{th}$ cell of the unstructured grid, and $U_i$, $p_i$ and $\nu_{ti}$, are the values of the velocity, pressure and eddy viscosity at this $i^{th}$ cell, respectively. The optimization algorithm, ADAM (Adaptive Moment Estimation) \cite{KingmaBa2015}, was selected for its efficiency and convergence properties. Our optimization algorithm is presented below.

\begin{algorithm}[htbp]
\caption{\textbf{Pure NN Training Algorithm}}\label{alg:purenn}
\begin{algorithmic}[1]
\Require Mini-batched velocity, pressure, and eddy viscosity data 
\[
\tilde{\boldsymbol{\phi}} = \{\tilde{p},\tilde{u},\tilde{v},\tilde{w},\tilde{\nu_t}\}
\]
along with their corresponding spatial and angular data
\[
\{x,y,z,\cos(\theta), \sin(\theta)\}
\]
\Require Initialize neural network parameters 
\(\boldsymbol{\theta}_{\tilde{\boldsymbol{\phi}}'} = \{\mathbf{W}^{(l)}, \mathbf{b}^{(l)}\}_{l=1}^{L}\)
\State Set hyperparameters (learning rate \(\eta\), batch size \(M\), maximum epochs \(S\))
\State Set iteration counter \(s = 0\)
\While{not converged (until loss reaches a threshold or max epochs \(S\))}
  \State Randomly sample a mini-batch of \(M\) data points:
  \[
  \{(x_i, y_i, z_i, \cos(\theta_i), \sin(\theta_i))\}_{i=1}^{M}
  \]
  \State Retrieve corresponding target values:
  \[
  \{ (\tilde{p}_i, \tilde{u}_i, \tilde{v}_i, \tilde{w}_i, \tilde{\nu}_{t_i}) \}_{i=1}^{M}
  \]
  \State \textbf{Forward pass:} Compute predictions \(\hat{\tilde{\boldsymbol{\phi}}}\) using the neural network.
  \State \textbf{Compute loss:} Evaluate the loss function \(\mathcal{L}_{\text{fit}}\)
  \State \textbf{Compute gradients:} Compute \(\partial_{\boldsymbol{\theta}_{\tilde{\boldsymbol{\phi}}'}} L_{\text{fit}}\).
  \State \textbf{Update parameters:} Using gradient-based optimization
  \[
  \boldsymbol{\theta}_{\tilde{\boldsymbol{\phi}}'}^{(s+1)} = \boldsymbol{\theta}_{\tilde{\boldsymbol{\phi}}'}^{(s)} - \eta \nabla_{\boldsymbol{\theta}_{\tilde{\boldsymbol{\phi}}'}} L_{\text{fit}}
  \]
  \State Increment iteration counter: \(s \gets s + 1\)
\EndWhile
\State \textbf{Output:} Optimized network parameters 
\[
\{\boldsymbol{\theta}_{\tilde{\boldsymbol{\phi}}'}^{(1)}, \boldsymbol{\theta}_{\tilde{\boldsymbol{\phi}}'}^{(2)}, \dots, \boldsymbol{\theta}_{\tilde{\boldsymbol{\phi}}'}^{(S)}\}
\]

\end{algorithmic}
\end{algorithm}

 \textbf{Hyper-parameters used in this paper:} $S = 30000$, $M = 2^{15}, \eta = 10^{-3}$.

\subsubsection{Results}

We employ two key evaluation metrics: (i) Mean Squared Error (MSE) to quantify the deviation of model predictions from CFD data, and (ii) R\(^2\) score to measure how well the regressor explains the variance in the CFD data by quantifying the proportion of total variance that it captures. These are presented in Table \ref{tab:purennresults} below. We evaluate the model on two angles not included in training: the intermediate angles $7.5^{\circ}$ and $157.5^{\circ}$ to assess interpolation capability between trained angles.

\begin{table}[htbp]
\centering
\caption{Performance Metrics Across Wind Angles $7.5^\circ$ and $157.5^\circ$}
\label{tab:purennresults}
\begin{tabular}{llcccc}
\hline
\textbf{Angle} & \textbf{Variable} & \textbf{MSE} & \textbf{RMSE} & \textbf{MAE} & \textbf{R\textsuperscript{2}} \\
\hline
\multirow{5}{*}{$7.5^\circ$}
 & $p$ & 0.014 & 0.119 & 0.056 & 0.991 \\
 & $u_x$ & 0.002 & 0.047 & 0.032 & 0.973 \\
 & $u_y$ & 0.004 & 0.063 & 0.038 & 0.998 \\
 & $u_z$ & 0.000 & 0.012 & 0.006 & 0.995 \\
 & $\nu_t$ & 0.068 & 0.260 & 0.171 & 0.999 \\
\hline
\multirow{5}{*}{$157.5^\circ$}
 & $p$ & 0.001 & 0.038 & 0.022 & 0.999 \\
 & $u_x$ & 0.001 & 0.034 & 0.025 & 0.997 \\
 & $u_y$ & 0.002 & 0.047 & 0.028 & 0.999 \\
 & $u_z$ & 0.000 & 0.009 & 0.004 & 0.998 \\
 & $\nu_t$ & 0.040 & 0.201 & 0.146 & 1.000 \\
\hline
\end{tabular}
\end{table}

\begin{figure}[htbp]
  \centering

  \begin{subfigure}[b]{0.85\textwidth}
    \centering
    \includegraphics[width=\textwidth]{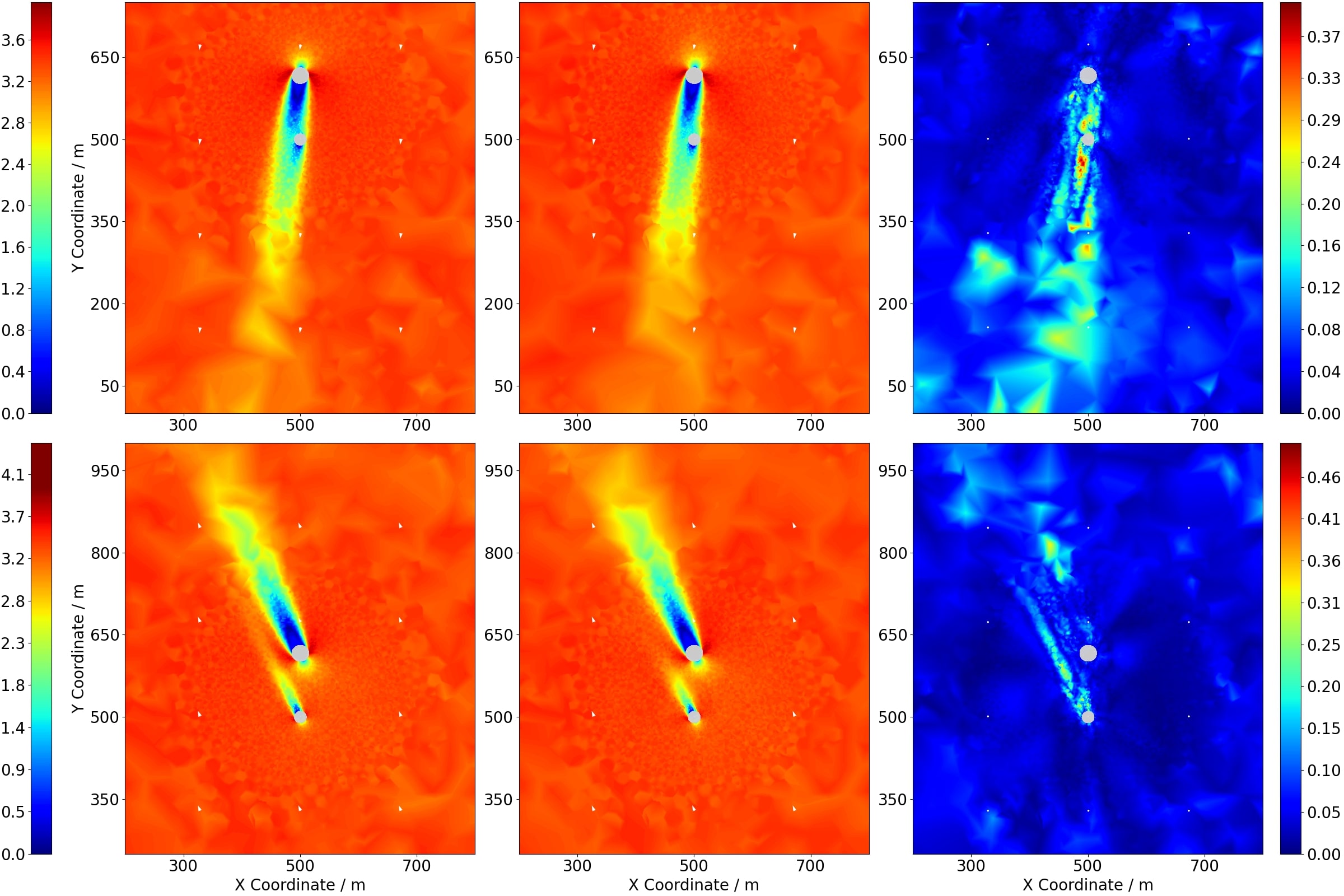}
    \caption{Velocity [$\mathrm{m}\,\mathrm{s}^{-1}$]}
    \label{fig:sub_total_velocity}
  \end{subfigure}
  \hfill
  \begin{subfigure}[b]{0.85\textwidth}
    \centering
    \includegraphics[width=\textwidth]{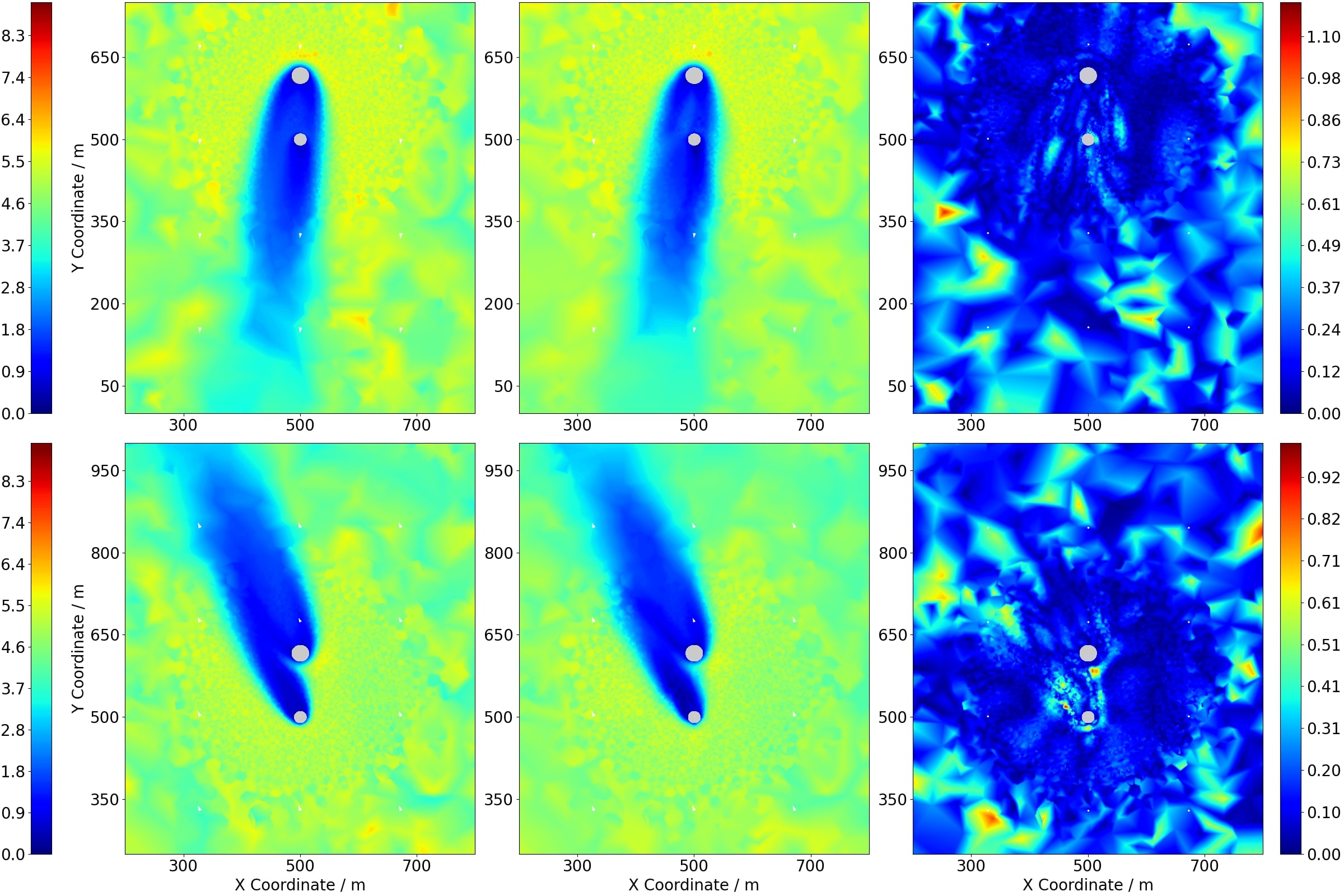}
    \caption{Eddy viscosity [$\mathrm{m}^2\mathrm{s}^{-1}$]}
    \label{fig:sub_turbvisc}
  \end{subfigure}

  \caption{Regression results for RANS fields using a fully connected neural network, shown at $Z=50\,\mathrm{m}$. Each image includes two rows corresponding to wind angles $7.5^\circ$ and $157.5^\circ$, and three columns showing ground truth (left), prediction (middle), and absolute error (right).}
  \label{fig:combined_nn_regression}
\end{figure}

For a visual validation, we show plots of the X-Y plane with a slice at the center of the sphere in Fig.~\ref{fig:combined_nn_regression}. We omit pressure as it bears the same conclusions as the velocity magnitude and eddy viscosity.






Although divergence was not explicitly constrained in our system, it was inherently satisfied since the data itself were divergence-free, within a tolerance determined by the spatial discretization error. The neural network results confirm this property, maintaining consistency with the data and preserving divergence-free behavior. In particular, the effect of the network smoothing on the velocity field results in divergence values that are even closer to zero compared to the original CFD data.

\subsubsection{Discussion}

\begin{figure}[htbp]
  \centering
  \includegraphics[width=1\textwidth]{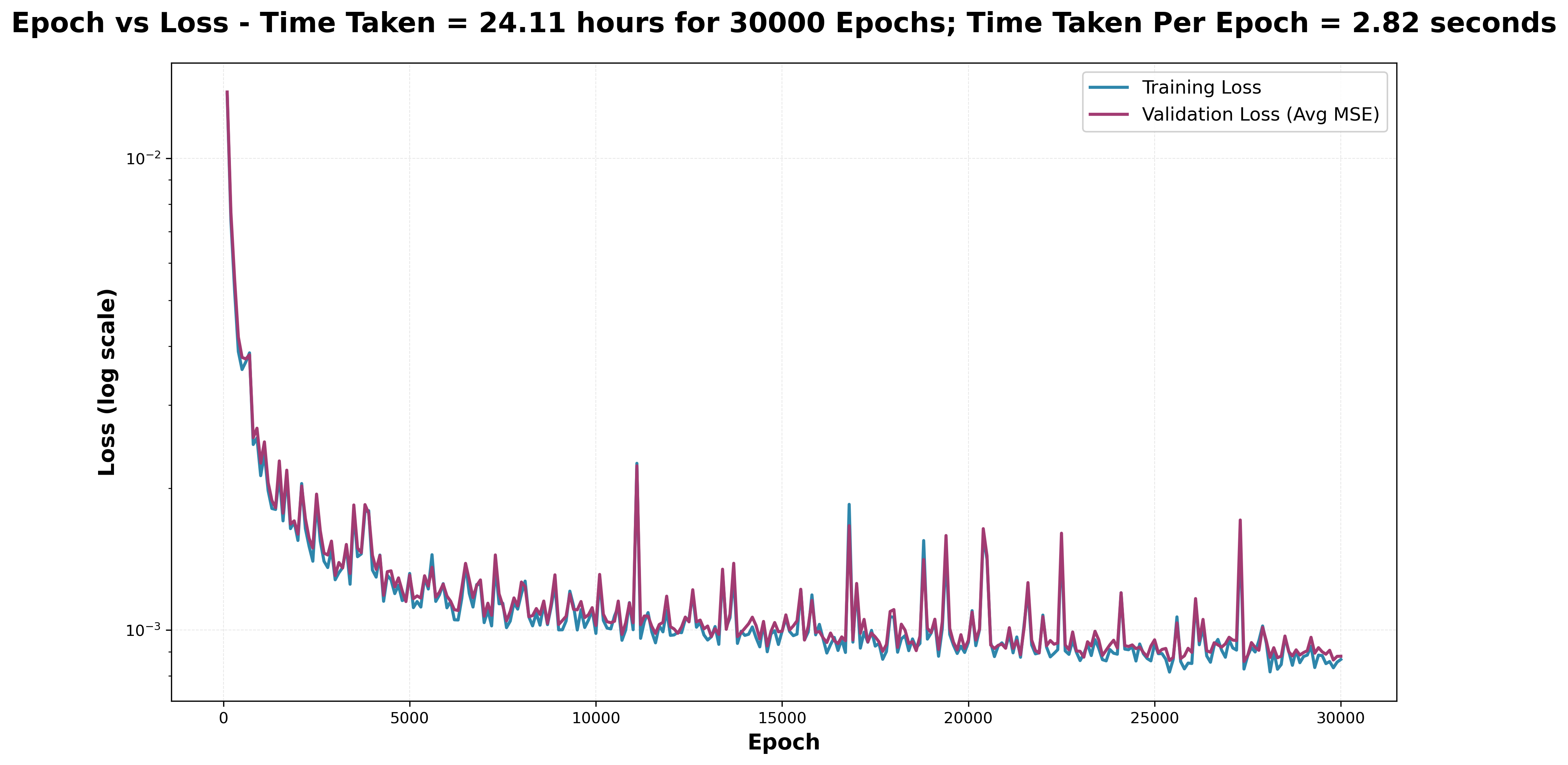}
  \caption{Training and validation loss curves for the pure neural network model over 30,000 epochs on NVIDIA RTX 3080 GPU. Y-axis shows log-scale MSE loss. Each epoch required approximately 2.82 seconds. Convergence is achieved after 30,000 epochs, reflecting the higher computational cost of the pure NN benchmark model.}
  \label{fig:save_logging_plot_ylog}
\end{figure}

The regression results align exceptionally well with the evaluation angles (intermediate interpolation points), establishing a strong accuracy benchmark for comparison. However, the computational cost is significant: on average, each epoch required approximately 2.82 seconds, running on a laptop equipped with a 16GB NVIDIA GeForce RTX 3080 GPU. The corresponding loss curve is shown in Fig.~\ref{fig:save_logging_plot_ylog}. We treat this pure NN as the reference benchmark for all comparisons; although more computationally demanding, it remains a fully feasible baseline.

We also observe that increasing network depth while reducing the number of hidden neurons significantly hinders convergence, making training more challenging. In contrast, a shallower network with more hidden neurons, while potentially trainable, would still demand similar or even greater computational time depending on the specific architecture. Finally, reducing the number of hidden neurons in the current network completely prevents convergence while increasing the number of hidden neurons encourages overfitting.

Despite the high accuracy achieved by this benchmark model, the prolonged training time motivates the development of more efficient alternatives for real-world applications. Scaling this direct approach to larger domains, such as full urban districts, would demand substantial computational resources. While this cost could in principle be addressed by upgrading to more powerful computing systems, we instead pursue a hybrid approach that leverages structured decomposition to achieve comparable accuracy with dramatically reduced training time, as outlined in the next section. Accordingly, the hybrid in Section~\ref{sec:hybridnn} is positioned as an accelerated approximate alternative evaluated against the Section~\ref{sec:purenn} benchmark.

\subsection{Hybrid Neural Interpolation with Data Decomposition}\label{sec:hybridnn}

In order to analyze and reduce the complexity of the RANS data under consideration, we propose a two-step method. In a first step, we decompose the high-dimensional data in a way accessible to a multilinear analysis, and in a second step, we introduce an NN correction to capture nonlinearities. The complete data set made of the 5 normalized scalar field values corresponding to the velocity components, pressure, and eddy viscosity on the unstructured grid at the 24 inlet angles sampled in increments of $15^{\circ}$ over the interval $[0^{\circ},345^{\circ}]$, is stored in the third order tensor $\mathcal{X} \in \mathbb{R}^{N \times M \times 5}$, where $N$ is the number of nodal points. Note that we sample up to $345^{\circ}$ to avoid duplication, as $360^{\circ}$ is equivalent to $0^{\circ}$ due to periodicity.

This method is approximate by design: it compresses the dataset via Tucker decomposition and learns residual corrections, trading some model flexibility for speed; we evaluate it against the pure NN benchmark in Section~\ref{sec:purenn}.

Code\_Saturne produces cell-centered fields for pressure, velocity, and eddy viscosity. In this subsection (unlike Section \ref{sec:purenn}), prior to visualization and post-processing, we interpolate cell-centered fields to nodal values (point values) using ParaView's built-in filter ('Cell Data to Point Data'), which computes vertex values as volume-weighted averages of adjacent cells. This pre-processing step eliminates piecewise-constant discontinuities that would otherwise corrupt any orthogonal decomposition. This also has the effect of improving training stability.

The number of nodal points for the model is $N = 100680$, about a third of the number of cells. For this subsection, we define $N\equiv100680$, $M\equiv24$.

\subsubsection{Tensor Tucker Decomposition}

Tucker decomposition \cite{Kolda2009,hitchcock1927,carroll1970} expresses a tensor of n-order in terms of a smaller core tensor and n orthonormal factor matrices. It is a multilinear generalization of Singular Value Decomposition (SVD) \cite{eckart1936} for matrices, with its higher order formulation - HOSVD - developed by De Lathauwer et al.~\cite{DeLathauwer2000}. For the three-order data tensor \(\mathcal{X}\), the Tucker decomposition writes as: 
\begin{equation}
\label{eq:tuckerein}
\mathcal{X} \approx \mathcal{G} \times_1 \mathbf{U} \times_2 \mathbf{V} \times_3 \mathbf{W}
\quad \Leftrightarrow \quad
\mathcal{X}_{ijk} \approx \mathcal{G}_{mnl} \, \mathbf{U}_{im} \, \mathbf{V}_{jn} \, \mathbf{W}_{kl},
\end{equation}

where \(\times_n\) denotes the \(n\)-mode product and repeated indices imply summation. The dimensions are: $i \in \{1,\ldots,N\}$ (spatial points), $j \in \{1,\ldots,M\}$ (wind angles), and $k \in \{1,\ldots,5\}$ (flow variables), while the core tensor indices are $m \in \{1,\ldots,120\}$, $n \in \{1,\ldots,24\}$, and $l \in \{1,\ldots,5\}$.

The core tensor $\mathcal{G} \in \mathbb{R}^{120 \times 24 \times 5}$ captures the interactions among the spatial, angular, and physical variable modes. The matrix $\mathbf{U} \in \mathbb{R}^{N \times 120}$ represents orthonormal spatial basis functions corresponding to nodal patterns in the mesh. The matrix $\mathbf{V} \in \mathbb{R}^{24 \times 24}$ spans the inlet wind angle mode,  each column captures a distinct angular variation in the velocity field. Lastly, $\mathbf{W} \in \mathbb{R}^{5 \times 5}$ encodes orthonormal interactions between physical variables (velocity components, pressure, eddy viscosity).

We deliberately avoided rank reduction in $\mathbf{V}$ and $\mathbf{W}$ to preserve full information in the angular and physical variable modes, which is essential for accurate interpolation across inlet angles and coupling between flow variables. Rank truncation is applied only along the spatial mode, with $\mathbf{U} \in \mathbb{R}^{N \times 120}$ obtained via HOSVD. The spatial rank of 120 is automatically determined by the HOSVD algorithm based on the data's intrinsic dimensionality, capturing the dominant spatial patterns while achieving significant compression from $N \approx 100{,}680$ nodal points. This is implemented via \textbf{TensorLy}'s \texttt{tucker} decomposition \cite{Kossaifi2019}.

To predict the velocity field at a new inlet wind angle $\theta_t$, we estimate a new row $V'(\theta_t) \in \mathbb{R}^{1 \times 24}$ and construct a zeroth-order approximation to the output tensor $\mathcal{X}_t^0 \in \mathbb{R}^{N \times 1 \times 5}$:
\begin{equation}
\label{eq:zeroth}
\mathcal{X}^0(\theta_t) \approx \mathcal{G} \times_1 \mathbf{U} \times_2 V'(\theta_t) \times_3 \mathbf{W}.
\end{equation}

This reconstruction forms the basis for subsequent correction steps.

\subsubsection{Angular Interpolation via $\textbf{V}$ with Fourier Series}\label{sec:vinterp}

We seek to construct a continuous periodic interpolation for each column (mode) of the matrix $V$ using Fourier series. Given $M=24$ wind angles uniformly spaced over $360^\circ$, we denote these angles as:
\begin{equation}
\theta_j = (j-1)\Delta, \quad \text{where} \quad \Delta = \frac{2\pi}{M} = 15^\circ, \quad j=1,\ldots,M.
\end{equation}

Our goal is to define, for each mode $m$, a $2\pi$-periodic continuous function $v^{(m)}(\theta)$ that interpolates the discrete values $V_j^{(m)} = V(\theta_j)$ at the given angles. We express it as a truncated Fourier series with $M$ modes:
\begin{equation}
\label{eq:fourieransatz}
v^{(m)}(\theta) = \Re\left\{\frac{1}{M}\sum_{k=1}^{M}\hat{V}^{(m)}_k e^{i(k-1)\theta}\right\}, \quad m=1,\ldots,M,
\end{equation}

To ensure exact interpolation at the training points, we require $v^{(m)}(\theta_j) = V_j^{(m)}$ for all $j$. This constraint determines the Fourier coefficients via the discrete Fourier transform:
\begin{equation}
\label{eq:vparam}
\hat{V}^{(m)}_k = \sum_{j=1}^{M} V_j^{(m)} \exp\left(-i\frac{2\pi(k-1)(j-1)}{M}\right) = \sum_{j=1}^{M} V_j^{(m)} \exp(-i(k-1)\theta_j).
\end{equation}

to obtain 
\begin{equation}
\label{eq:vparam2}
v^{(m)}(\theta) = \Re\left\{\frac{1}{M}\sum_{k=1}^{M}\sum_{j=1}^{M}V_j^{(m)}e^{i(k-1)(\theta-\theta_j)}\right\}
\end{equation}

and after some standard algebra, the interpolant for mode $m$ can be written compactly as:

\begin{equation}
\label{eq:fourieransatzredux}
v^{(m)}(\theta) = \frac{1}{M}\sum_{j=1}^M V_j^{(m)} K_j(\theta),
\end{equation}

where $K_j(\theta)$ is defined as 

\begin{equation}
\label{eq:interpkernel}
K_j(\theta) = \frac{1}{2}\left(1+\ \frac{\sin\left(M (\theta-\theta_j) - \frac{\theta-\theta_j}{2}\right)}{\sin\left(\frac{\theta-\theta_j}{2}\right)}\right),
\end{equation}
which can also be expressed in terms of the Dirichlet kernel, $D_{M-1}$
\begin{equation}
\label{eq:interpkernelDirich}
K_j(\theta) = \frac{1}{2}\left(1+D_{M-1}(\theta - \theta_j)\right).
\end{equation}

The kernel $K_j(\theta)$ ensures exact interpolation at the training points while providing smooth interpolation between them. Specifically, 
\begin{equation}
K_j(\theta_i) = M\delta_{ij}.
\end{equation}
Note that for any angle $\theta = \theta_j \pm (n+1/2)\Delta$, $D_{M-1}=1$ and $K_j=1$ too. Thus, for such an angle that is equidistant from two successive training angles, all components $V_j^m$ contribute equally to the $v^m$ mode.
Collecting all modes together, we obtain the full vector-valued interpolant:
\begin{equation}
V'(\theta_t) = \left[v^{(1)}(\theta_t), v^{(2)}(\theta_t), \ldots, v^{(M)}(\theta_t)\right],
\end{equation}

which provides a continuous approximation of the mode vector for any wind angle $\theta_t$.

To assess interpolation performance, we compare predictions at $7.5^{\circ}$ and $157.5^{\circ}$, with true data, both excluded from the training set. The prediction accuracy is quantified using four metrics: Mean Squared Error (MSE), Root Mean Squared Error (RMSE), Mean Absolute Error (MAE), and the coefficient of determination ($R^2$), as introduced in Section~\ref{sec:purenn}. Table~\ref{tab:tuckermetrics} summarizes the results for each flow variable at the test angles. The reconstruction demonstrates high fidelity, successfully capturing complex flow features with $R^2$ values exceeding 0.99 in most cases.

\begin{table}[htbp]
\centering
\caption{Performance Metrics for Tucker Decomposition and Fourier Interpolation at Test Angles $7.5^{\circ}$ and $157.5^{\circ}$}
\label{tab:tuckermetrics}
\begin{tabular}{llcccc}
\hline
\textbf{Angle} & \textbf{Variable} & \textbf{MSE} & \textbf{RMSE} & \textbf{MAE} & \textbf{R\textsuperscript{2}} \\
\hline
\multirow{5}{*}{$7.5^{\circ}$}
 & $p$ & 0.025 & 0.158 & 0.084 & 0.991 \\
 & $u_x$ & 0.004 & 0.063 & 0.033 & 0.967 \\
 & $u_y$ & 0.018 & 0.135 & 0.072 & 0.993 \\
 & $u_z$ & 0.001 & 0.038 & 0.015 & 0.971 \\
 & $\nu_t$ & 0.058 & 0.241 & 0.128 & 0.999 \\
\hline
\multirow{5}{*}{$157.5^{\circ}$}
 & $p$ & 0.006 & 0.079 & 0.039 & 0.998 \\
 & $u_x$ & 0.005 & 0.069 & 0.030 & 0.990 \\
 & $u_y$ & 0.015 & 0.121 & 0.056 & 0.994 \\
 & $u_z$ & 0.001 & 0.037 & 0.013 & 0.973 \\
 & $\nu_t$ & 0.021 & 0.144 & 0.090 & 1.000 \\
\hline
\end{tabular}
\end{table}

\begin{figure}[htbp]
  \centering

  \begin{subfigure}[b]{0.85\textwidth}
    \includegraphics[width=\textwidth]{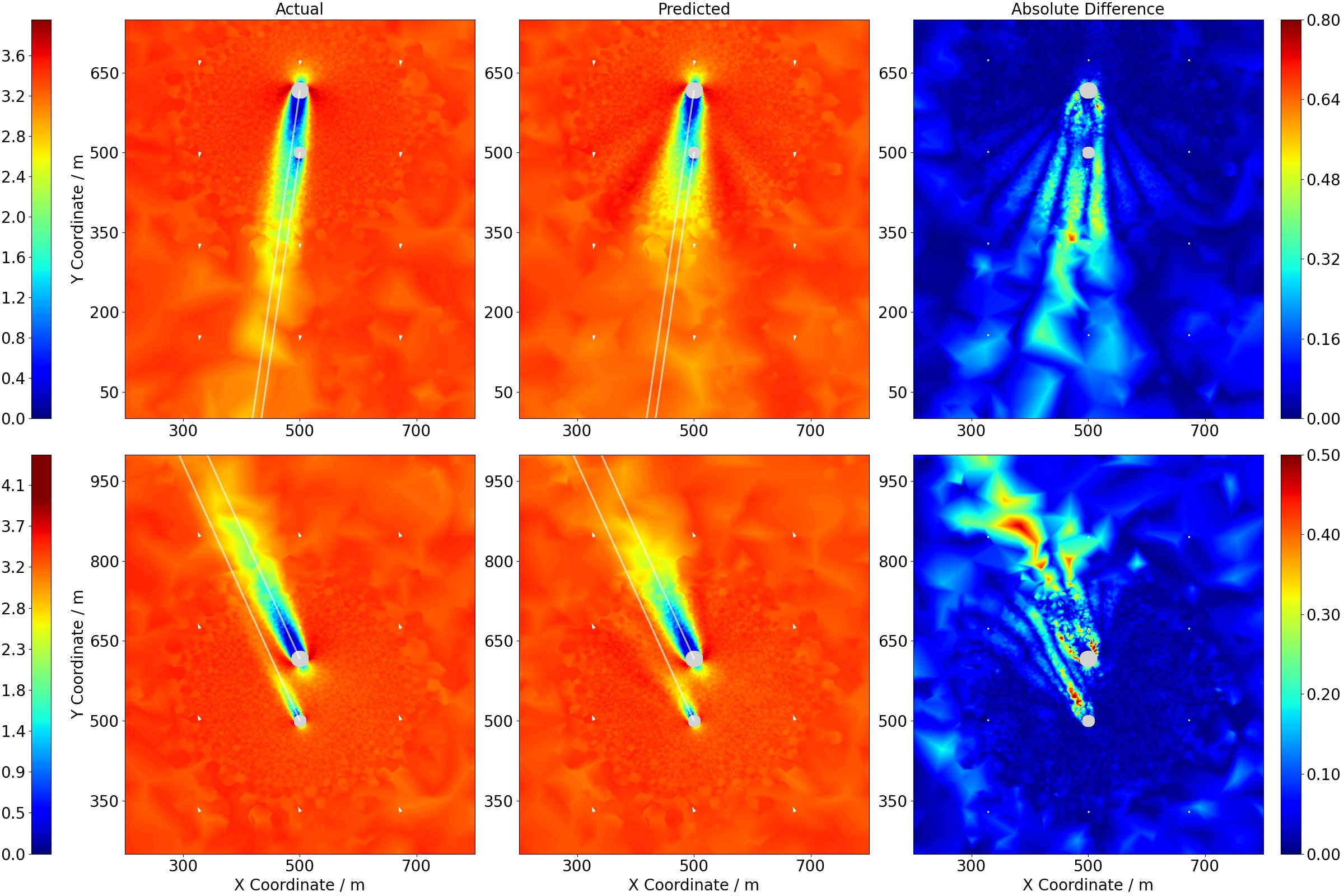}
    \caption{Velocity [$\mathrm{m}\,\mathrm{s}^{-1}$]}
    \label{fig:velocity_init}
  \end{subfigure}
  \begin{subfigure}[b]{0.85\textwidth}
    \includegraphics[width=\textwidth]{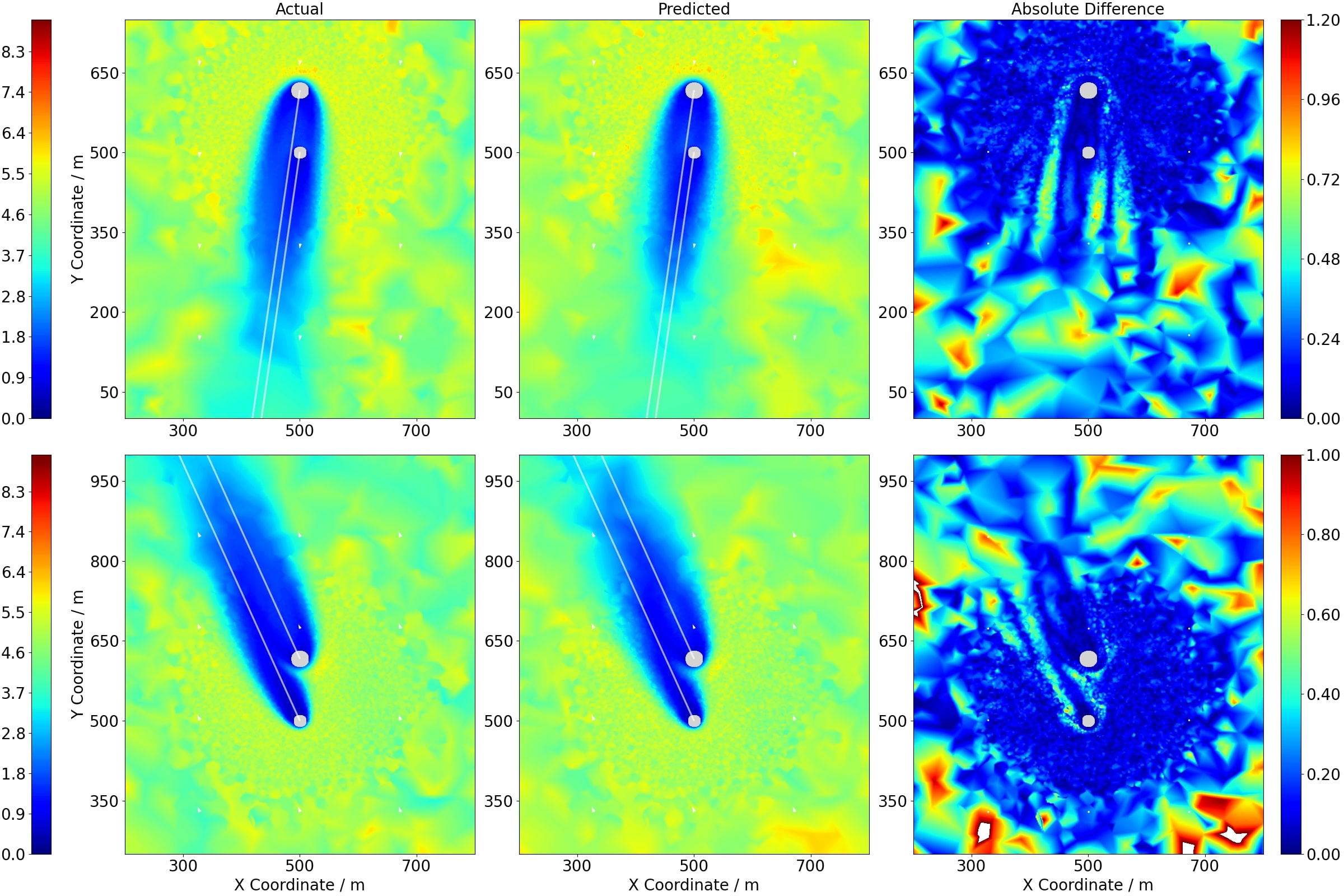}
    \caption{Eddy viscosity [$\mathrm{m}^2\mathrm{s}^{-1}$]}
    \label{fig:turbvisc_init}
  \end{subfigure}

  \caption{Tucker-Fourier interpolation of RANS fields at $Z=50\,\mathrm{m}$ for inlet angles $7.5^\circ$ (top row) and $157.5^\circ$ (bottom row). Each subfigure shows the ground truth (left), Tucker-Fourier prediction (middle), and absolute error (right) for velocity, eddy viscosity, and pressure fields.}
  \label{fig:tucker_init}
\end{figure}

For a visual assessment, the velocity magnitude, eddy viscosity and pressure in the X-Y plane at $Z = 50$ m is shown in Fig.~\ref{fig:tucker_init}. 

Although the metrics suggest an accurate recovery of flow variables, the visualizations of the predicted fields reveal structural artifacts, notably a ringing effect near discontinuities, evident in Fig.~\ref{fig:tucker_init}. For the two test angles, $7.5^{\circ}$ and $157.5^{\circ}$, we attribute these artifacts to the interpolation kernel defined in Eq.~\ref{eq:interpkernel}, which is based on Fourier modes. In particular, the issue arises because the RANS datasets are discontinuous across angles - each simulation corresponds to a distinct set of boundary conditions. Fourier interpolation, relying on smooth and globally supported sinusoids, struggles to represent such discontinuities using a finite number of modes. This mismatch results in localized overshoots and undershoots near angular transitions, a manifestation of the Gibbs phenomenon \cite{gibbs1899fourier,gottlieb1997,hewitt1979}. These observations motivate a nonlinear neural correction strategy in which a neural network (NN) is used to mitigate the limitations of global interpolation kernels and enhance generalization across the angular domain. This is presented below.

\subsubsection{Spatial Interpolation via $\textbf{U}$ with \(k\)-Nearest Neighbors}

To interpolate the factor matrix $\textbf{U}$ with respect to the spatial Cartesian coordinate $R$, we proceed as follows. Given \(N\) spatial points \(\{c_i\}_{i=1}^N\subset\mathbb{R}^3\) (training centers) with associated factor values \(\{u_i\}_{i=1}^N\subset\mathbb{R}^{120}\) from the matrix $\mathbf{U}$, we employ a k-nearest neighbors approach \cite{fix1951,cover1967} with Gaussian radial basis function kernels \cite{Buhmann2003,Wendland2004}. We fix a Gaussian kernel width \(\sigma>0\) and a neighbour count \(k\). For an arbitrary point \(R\in\mathbb{R}^3\), we first compute the squared Euclidean distances,
\begin{equation}
d_i^2(R)=\|R - c_i\|^2,\quad i=1,\dots,N,
\end{equation}
and let \(I_k(R)\subset\{1,\dots,N\}\) denote the indices of the \(k\) smallest values among \(\{d_i^2(R)\}\) \cite{Fasshauer2007}. We then assign each neighbor \(i\in I_k(R)\) a Gaussian weight,
\begin{equation}
w_i(R)
\;=\;
\exp\!\biggl(-\frac{d_i^2(R)}{2\,\sigma^2}\biggr)\,.
\end{equation}

and form the normalized interpolant
\begin{equation}
\label{eq:uparam}
\widehat{U}(R)
=\frac{\sum_{i\in I_k(R)}w_i(R)\,u_i}
   {\sum_{i\in I_k(R)}w_i(R)}\,.
\end{equation}

For a batch of \(N\) positions \(\{R_j\}_{j=1}^N\), the same formulas apply independently to each \(R_j\). The kernel width \(\sigma\) and neighbor count \(k\) were selected to balance interpolation smoothness with local feature preservation. We first analyzed the characteristic spacing of the unstructured grid, finding a median nearest-neighbor distance of \(h \approx 3.41\), and set the Gaussian kernel width as \(\sigma = 2h \approx 6.82\), following the principle that the kernel support (effective radius \(\approx 3\sigma\)) should encompass multiple grid points to ensure smooth reconstruction. With this kernel width, we selected \(k = 15\) neighbors to adequately sample the Gaussian weight distribution while maintaining computational efficiency. Cross-validation on the Tucker factor matrices confirmed this configuration minimizes reconstruction error (RMSE \(<\) 0.01) for the training angles while preserving the smoothness properties of the CFD data.

\subsubsection{Zeroth Order Ansatz}

With the parameterization of the factor matrix $U$ and $V$ given by Eqs.~\eqref{eq:uparam} and \eqref{eq:vparam} (re-expressed in kernel form as per Eq.~\eqref{eq:interpkernel}), we rewrite the zeroth-order solution of the parameterized Tucker Decomposition or zeroth-order ansatz given in Eq.~\eqref{eq:zeroth} as 

\begin{equation}
\label{eq:zerothansatz}
\widehat{\mathcal{X}_k^0}(R, \theta) \;\approx\; G_{mnl} 
\;\Biggl\{\,
\frac{
 \displaystyle\sum_{i=1}^N w_i(R) \ U_{i m}
}{
 \displaystyle\sum_{i=1}^N w_i(R)
}
\Biggr\}
\;\;
\;\;
\Biggl\{
 \displaystyle\sum_{j=1}^M V_{j n}\,K_j(\theta)
\Biggr\}
\;\;
\Bigl\{W_{k l}\Bigr\}
\end{equation}

where we have utilized the tensor notation expressed in Eq.~\eqref{eq:tuckerein}, and the hat $\hat{.}$ represents our notation for a Tucker reconstructed tensor. 

\subsection{Enhanced Ansatz Embedding: Self-Consistent Correction and Structured Perturbations}

Our neural network architecture centers on embedding a physics-informed ansatz $\widehat{\mathcal{X}^0}(R,\theta)$ through a bottleneck layer. This design addresses two key challenges observed in initial results: improving solution fidelity and eliminating ringing artifacts (Fig.~\ref{fig:velocity_init}). In the hybrid network we use the Rectified Linear Unit (ReLU) activation \cite{Glorot2011,nair2010} to promote sparse, nonnegative residual corrections and stable gradients in the bottleneck path:

\begin{equation}
f_{\mathrm{ReLU}}(x) = \max(0, x).
\end{equation}

We enhance the embedded ansatz through two complementary mechanisms that work in tandem: (i) \emph{self-consistent correction} forces network predictions to converge toward known reference solutions by incorporating residual information from training data, while (ii) \emph{structured perturbations} promote robust generalization across angular variations by systematically modifying the input ansatz.

The correction mechanism integrates additively after the first hidden layer, creating a hybrid forward pass:
\begin{equation}
\label{eq:bottleneck_clean}
\begin{aligned}
Z^{(1)} &= W^{(1)} A^{(0)} + B^{(1)}, \\
A^{(1)} &= f_{\mathrm{ReLU}}\!\left(Z^{(1)}\right) 
    + f_{\mathrm{ReLU}}\!\Big(W^{(b)} \big(\widehat{\mathcal{X}}^0 + \widetilde{\delta \mathcal{X}}^{(t)}\big) + B^{(b)}\Big), \\
Z^{(j)} &= W^{(j)} A^{(j-1)} + B^{(j)}, \quad &j = 2, \dots, \numlayers, \\
A^{(j)} &= f_{\mathrm{ReLU}}\!\left(Z^{(j)}\right), \quad &j = 2, \dots, \numlayers, \\
Y^{(i)} &= W^{(o,i)} A^{(\numlayers)} + B^{(o,i)}, \quad &i = 1, \dots, 5.
\end{aligned}
\end{equation}
Our main hypothesis lies in the second term of $A^{(1)}$, where the corrected ansatz $\widehat{\mathcal{X}}^0 + \widetilde{\delta \mathcal{X}}^{(t)}$ enters through dedicated bottleneck weights $W^{(b)}$.

At each training epoch $t$, we compute a correction tensor that captures the discrepancy between network predictions and reference data:
\begin{equation}
\label{eq:sc_corr}
\widetilde{\delta \mathcal{X}}_k^{(t)}(R_i, \theta) 
= \sum_{j=1}^{M} \Big(\mathcal{X}_{ijk} - \tilde{\mathcal{X}}_k^{(t-1)}(R_i, \theta_j)\Big)\, \alpha_j(\theta),
\end{equation}
where $\mathcal{X}_{ijk}$ represents the reference solution, $\tilde{\mathcal{X}}_k^{(t-1)}$ is the network's prediction from the previous epoch, and $\alpha_j(\theta)$ are normalized interpolation weights.

The interpolation weights derive from a correction kernel that naturally suppresses high-frequency oscillations:
\begin{equation}
\label{eq:fejernn}
\tilde{K}_j(\theta) = \frac{1}{2}\left\{ 1 + \frac{\sin^2\left( N\frac{\theta - \theta_j}{2}\right)}{N\sin^2\left(\frac{\theta - \theta_j}{2}\right)} \right\}
\end{equation}
where we have set $N = M -1$, with $\alpha_j(\theta) = \tilde{K}_j(\theta)/\sum_{l=1}^{M} \tilde{K}_l(\theta)$ and normalization ensuring $\sum_{j=1}^{M} \alpha_j(\theta) = 1$. This kernel emerges from averaging the partial sums in Eq.~\eqref{eq:interpkernelDirich} (detailed in~\ref{sec:fejer}) and was scaled such that its product with Eq.~\eqref{eq:interpkernel} matches the peak of Eq.~\eqref{eq:interpkernel}. Eq.~\eqref{eq:fejernn} provides natural regularization by attenuating high-frequency modes that typically cause numerical instabilities near discontinuities.

The correction mechanism exhibits desirable convergence behavior: when network predictions $\tilde{\mathcal{X}}_k^{(t-1)}$ match reference data $\mathcal{X}_{ijk}$ across all angular samples, the residuals vanish identically, yielding $\widetilde{\delta \mathcal{X}}^{(t)} \equiv 0$. This means the correction acts purely as a transient training signal - once the network learns the target mapping, the architecture reduces to standard feedforward operation. Training begins with $\tilde{\mathcal{X}}^{(-1)} \equiv \widehat{\mathcal{X}^0}$, using the initial ansatz as the baseline prediction.

The self-consistent correction can be
viewed as a Picard iteration scheme \cite{picard1893,arnold1992ode}, a classical fixed-point iteration method \cite{banach1922,granas2003fixed}. Define the operator $T[\tilde{\mathcal{X}}] =
\text{NN}(\widehat{\mathcal{X}^0} + \text{Interp}[\mathcal{X} - \tilde{\mathcal{X}}])$
where Interp[·] denotes angular interpolation. The fixed-point equation
$\tilde{\mathcal{X}} = T[\tilde{\mathcal{X}}]$ has solution $\tilde{\mathcal{X}} = \mathcal{X}$
when the network perfectly reconstructs training data.

\paragraph{Structured Perturbations for Angular Robustness}
To suppress spurious oscillations and improve generalization across unseen angles, 
we introduce structured perturbations that mimic angular artifacts. 
Given angular samples \(\{\theta_j\}_{j=1}^M\), we define neighbours
\[
\mathcal{N}_j = \{\theta_j + d \bmod 360^\circ \mid d \in D\}, \quad 
D = \{-\Delta, -\Delta/2, \Delta/2, \Delta\}, \quad \Delta=15^\circ.
\]
Perturbations are computed as
\[
\delta_{i,j}^{(d)} = \widehat{\mathcal{X}^0}(R_i, \theta_j+d) - \widehat{\mathcal{X}^0}(R_i, \theta_j), \quad d\in D,
\]
and weighted using a damped Dirichlet kernel, $D_{2M-2}$, as per Eq.~\eqref{eq:interpkernelDirich}:
\[
w_d = D_{2M-2}(\phi_d)\, e^{-0.1|\phi_d|}, \qquad
\phi_d = \tfrac{2\pi d}{360}.
\]
After normalization, $\bar{w}_d = w_d/\sum_{d' \in D}|w_{d'}|$, 
the perturbation-augmented ansatz is defined as
\begin{equation}
\label{eq:noisyaug}
\tilde{\mathcal{x}}_{i,j} = \widehat{\mathcal{X}^0}(R_i,\theta_j) + \sum_{d\in D} \bar{w}_d\, \delta_{i,j}^{(d)}.
\end{equation}

During training, the bottleneck input in Eq.~\eqref{eq:bottleneck_clean} is modified by replacing 
\(\widehat{\mathcal{X}^0}\) with \(\tilde{\mathcal{x}}\):
\begin{equation}
\label{eq:bottleneck_aug}
A^{(1)} = f_{\mathrm{ReLU}}\!\left(Z^{(1)}\right) 
    + f_{\mathrm{ReLU}}\!\Big(W^{(b)} \big(\tilde{\mathcal{x}} + \widetilde{\delta \mathcal{X}}^{(t)}\big) + B^{(b)}\Big).
\end{equation}

We rewrite Eq.~\eqref{eq:bottleneck_clean} with the addition of Eq.~\eqref{eq:bottleneck_aug}.

\begin{equation}
\label{eq:nnsolution}
\begin{aligned}
Z^{(1)} &= W^{(1)} A^{(0)} + B^{(1)}, \\
A^{(1)} &= f_{\mathrm{ReLU}}\!\left(Z^{(1)}\right) 
    + f_{\mathrm{ReLU}}\!\Big(W^{(b)} \big(\tilde{\mathcal{x}} + \widetilde{\delta \mathcal{X}}^{(t)}\big) + B^{(b)}\Big), \\
Z^{(j)} &= W^{(j)} A^{(j-1)} + B^{(j)}, \quad &j = 2, \dots, \numlayers, \\
A^{(j)} &= f_{\mathrm{ReLU}}\!\left(Z^{(j)}\right), \quad &j = 2, \dots, \numlayers, \\
Y^{(i)} &= W^{(o,i)} A^{(\numlayers)} + B^{(o,i)}, \quad &i = 1, \dots, 5.
\end{aligned}
\end{equation}

Convergence follows from the contraction property \cite{banach1922,granas2003fixed}: if the network is 
Lipschitz continuous with constant $L < 1$, then 
$\|\tilde{\mathcal{X}}^{(t+1)} - \tilde{\mathcal{X}}^{(t)}\| \leq L \|\tilde{\mathcal{X}}^{(t)} - \tilde{\mathcal{X}}^{(t-1)}\|$, 
ensuring convergence to the fixed point. The structured perturbations 
regularize this iteration by preventing over-fitting to interpolation 
artifacts.

\paragraph{Loss Function and Optimization Strategy}
The training objective combines mean squared error (MSE) and a differentiable $R^2$ loss~\cite{kumari2021r2}:
\begin{equation}
\label{eq:hybridloss}
\mathcal{L}_{\text{tot},k} =
\underbrace{\frac{1}{NM} \sum_{i=1}^{N} \sum_{j=1}^{M}
\left\| \tilde{\mathcal{X}}_{ijk} - \mathcal{X}_{ijk} \right\|^2}_{\mathcal{L}_{\text{MSE}}}
+
\underbrace{1 - \frac{\sum_{ij} \left\| \tilde{\mathcal{X}}_{ijk} - \mathcal{X}_{ijk} \right\|^2}
{\sum_{ij} \left\| \mathcal{X}_{ijk} - \bar{\mathcal{X}}_{jk} \right\|^2 + \varepsilon}}_{\mathcal{L}_{R^2}}.
\end{equation}

Here, $\tilde{\mathcal{X}}_{ijk}$ is the predicted output from Eq.~\eqref{eq:nnsolution} at spatial point $R_i$, wind angle $\theta_j$ and parameter $k$, while $\mathcal{X}_{ijk}$ denotes the corresponding ground truth data. The mean $\bar{\mathcal{X}}_{jk}$ is computed over all $i$ for each angle $\theta_j$ and parameter $k$, and $\varepsilon$ ensures numerical stability. The total loss is then,

\begin{equation}
\label{eq:totalloss}
\mathcal{L}_{\text{tot}} = \frac{1}{k}\sum_k\mathcal{L}_{\text{tot},k}
\end{equation}

The optimizer is Adam with a cosine warm-up over \( \warmupsteps \) steps and \texttt{StepLR} decay every \decayinterval{} steps. We clip gradients to a norm of \( \clipnorm \), initialize weights with identity maps (except input: orthogonal \cite{Saxe2014} with gain \( \initgain \)), and fix all biases to zero. Training is run for \totepochs{} epochs with checkpointing every \checkinterval{} epochs. The complete hybrid architecture contains 16,197 trainable parameters, representing a 3.1-fold reduction compared to the pure NN baseline (50,949 parameters).

\paragraph{Summary}
This hybrid approach embeds known physical priors via an ansatz, enforces angular consistency through augmentation, and iteratively refines corrections in a self-consistent fashion. The result is a robust, low-oscillation surrogate model that accurately predicts velocity, pressure, and eddy viscosity fields across varying inlet angles. The procedure is summarized in Algorithm~\ref{alg:hybridnn}.

\begin{algorithm}[htbp]
\caption{\textbf{Self-Consistent Residual Learning with Embedded Ansatz and Angular Augmentation}}
\label{alg:hybridnn}
\scalebox{0.78}{
\parbox{\linewidth}{
\begin{algorithmic}[1]
\Require Dataset $\{(R_i, \theta_j, \mathcal{X}_{ijk})\}$, Tucker components $(\mathcal{G}, \mathbf{V}, \mathbf{W})$, angular centers $\{(c_i, u_i \in \mathbf{U})\}$, shift range $s$, angular kernel $D_s(\phi)$, optimizer parameters $(\eta_{\min}, \eta_{\max}, T_w, \gamma, T_d, T)$, clip norm $c$, mode flag \texttt{is\_training}
\Ensure Final model parameters $\Theta^*$ (if training), or predictions $\tilde{\mathcal{X}}(R, \theta)$ (if inference)

\vspace{0.5em}
\If{\texttt{is\_training}}
  \State Initialize weights $\Theta$, set \( \tilde{\mathcal{X}}^{(-1)}(R, \theta) \coloneqq \widehat{\mathcal{X}^0}(R, \theta) \) using Eq.~\eqref{eq:zerothansatz}
  \For{$t = 1$ to $T$}
    \State Update learning rate via warm-up and decay schedule
    \State Compute residual correction $\widetilde{\delta \mathcal{X}}^{(t)}(R, \theta)$ using Eq.~\eqref{eq:sc_corr}
    \State Form embedded input: $\tilde{\mathcal{X}}^{(t)}(R, \theta) = \widehat{\mathcal{X}^0}(R, \theta) + \widetilde{\delta \mathcal{X}}^{(t)}(R, \theta)$
    \For{each training pair $(R_i, \theta_j)$}
      \State Generate angular-augmented target $\tilde{\mathcal{x}}_{i,j}$ using Eq.~\eqref{eq:noisyaug}
      \State Perform network forward pass using Eq.~\eqref{eq:nnsolution}, with embedded input $\tilde{\mathcal{X}}^{(t)}(R_i, \theta_j)$
    \EndFor
    \State Compute total loss $\mathcal{L}^{(t)}$, clip gradients to norm $c$, update weights $\Theta$
  \EndFor
  \State \Return $\Theta^*$
\Else
  \Comment{\textbf{Inference mode (no correction or augmentation)}}
  \For{each query $(R, \theta)$}
    \State Compute zeroth-order ansatz $\widehat{\mathcal{X}^0}(R, \theta)$ using Eq.~\eqref{eq:zerothansatz}
    \State Evaluate network prediction $\tilde{\mathcal{X}}(R, \theta)$ using Eq.~\eqref{eq:nnsolution} with $\widehat{\mathcal{X}^0}(R, \theta)$ only
    \State \Return $\tilde{\mathcal{X}}(R, \theta)$
  \EndFor
\EndIf
\end{algorithmic}
}}
\end{algorithm}

\subsubsection{Results}

We evaluate the performance of our hybrid Tucker-NN model using both quantitative metrics and visualizations at two held-out test angles, $7.5^\circ$ and $157.5^\circ$. Metrics used are consistent with those in Sections~\ref{sec:purenn} and~\ref{sec:vinterp}.

\paragraph{1. Quantitative Results}

Table~\ref{tab:hybrid_combined} reports the MSE, RMSE, MAE, and $R^2$ scores across the five predicted fields. Results confirm strong accuracy across all five fields at both test angles.

\begin{table}[htbp]
\centering
\caption{Performance metrics for the hybrid Tucker-NN model at test angles $7.5^\circ$ and $157.5^\circ$}
\label{tab:hybrid_combined}
\begin{tabular}{llcccc}
\hline
\textbf{Angle} & \textbf{Variable} & \textbf{MSE} & \textbf{RMSE} & \textbf{MAE} & \textbf{R\textsuperscript{2}} \\
\hline
\multirow{5}{*}{$7.5^\circ$}
 & $p$ & 0.031 & 0.175 & 0.099 & 0.989 \\
 & $u_x$ & 0.006 & 0.075 & 0.055 & 0.953 \\
 & $u_y$ & 0.015 & 0.122 & 0.080 & 0.994 \\
 & $u_z$ & 0.001 & 0.033 & 0.018 & 0.977 \\
 & $\nu_t$ & 0.060 & 0.244 & 0.136 & 1.000 \\
\hline
\multirow{5}{*}{$157.5^\circ$}
 & $p$ & 0.020 & 0.140 & 0.076 & 0.993 \\
 & $u_x$ & 0.004 & 0.067 & 0.044 & 0.990 \\
 & $u_y$ & 0.010 & 0.099 & 0.058 & 0.996 \\
 & $u_z$ & 0.001 & 0.030 & 0.016 & 0.982 \\
 & $\nu_t$ & 0.016 & 0.126 & 0.088 & 1.000 \\
\hline
\end{tabular}
\end{table}

\paragraph{2. Field Comparison Against CFD}

\begin{figure}[htbp]
  \centering

  \begin{subfigure}[b]{0.85\textwidth}
    \centering
    \includegraphics[width=\textwidth]{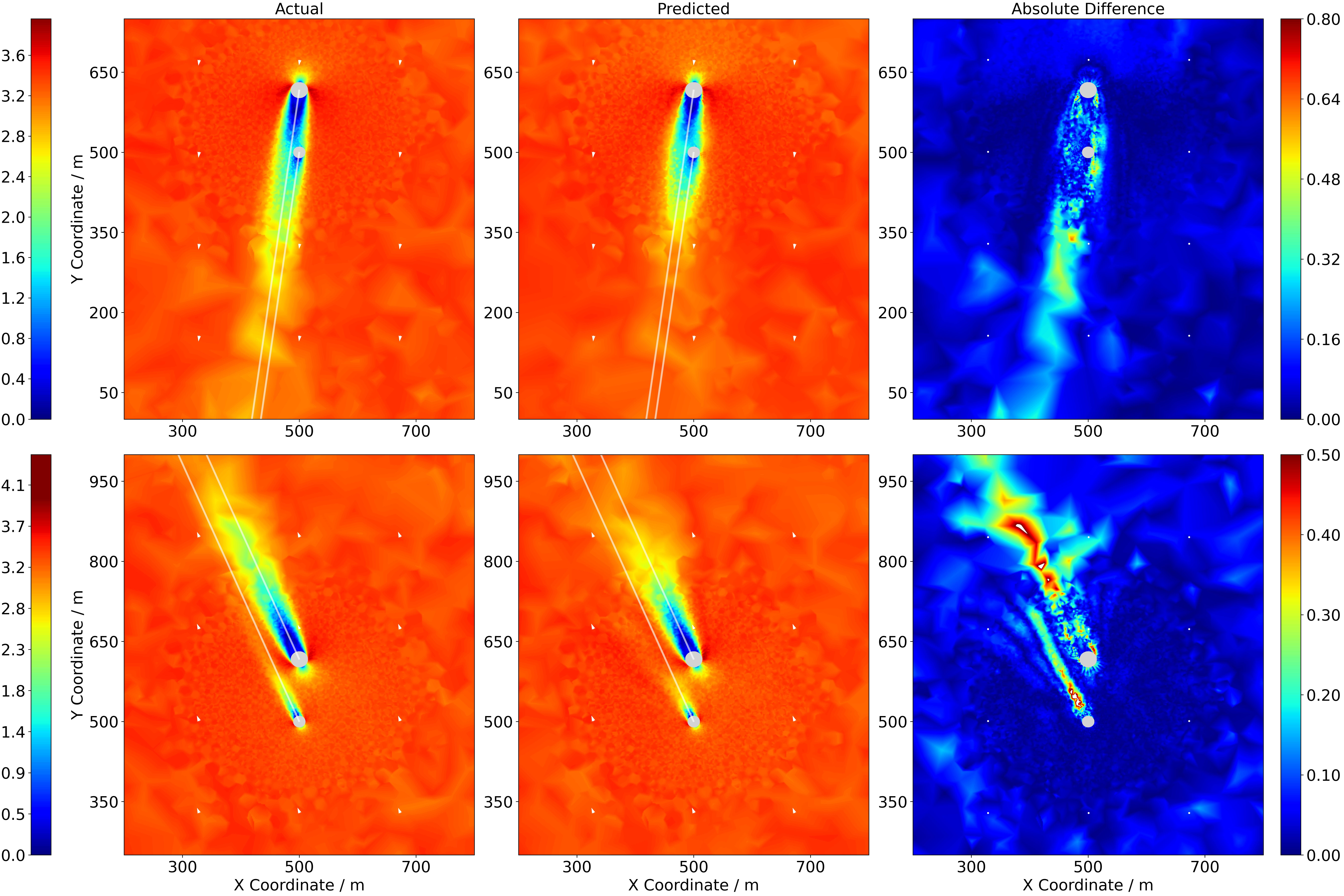}
    \caption{Velocity [$\mathrm{m}\,\mathrm{s}^{-1}$]}
    \label{fig:hy_velocity}
  \end{subfigure}
  \hfill
  \begin{subfigure}[b]{0.85\textwidth}
    \centering
    \includegraphics[width=\textwidth]{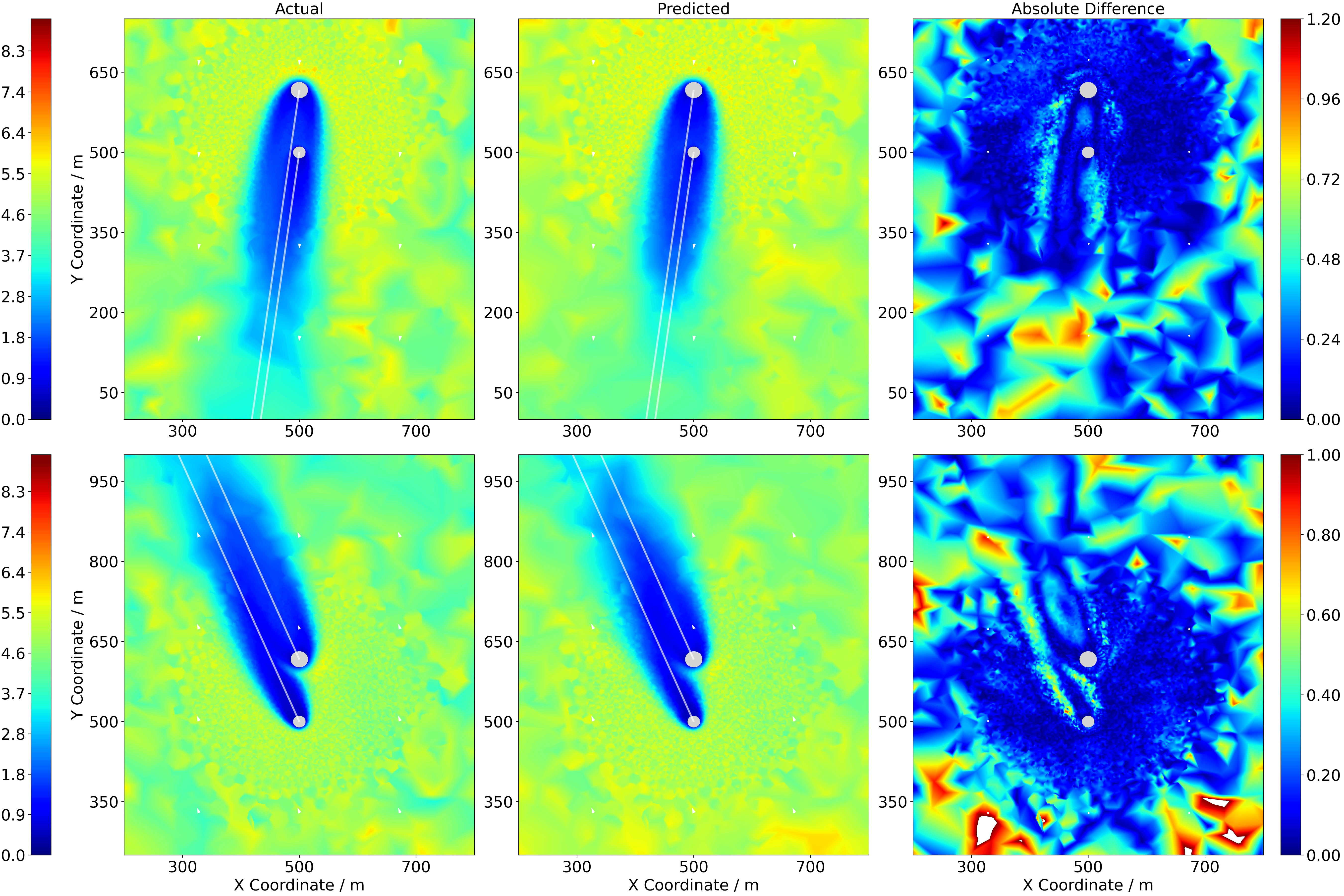}
    \caption{Eddy viscosity [$\mathrm{m}^2\,\mathrm{s}^{-1}$]}
    \label{fig:hy_turbvisc}
  \end{subfigure}

  \caption{Hybrid Tucker-NN model predictions for velocity magnitude, eddy viscosity, and pressure fields in the $Z = 50\,\mathrm{m}$ plane. Each subplot contains two rows for test angles $7.5^\circ$ (top) and $157.5^\circ$ (bottom), and three columns showing CFD ground truth (left), hybrid model prediction (middle), and absolute error (right). The hybrid correction notably improves the accuracy of eddy viscosity fields, especially near regions previously affected by ringing.}
  \label{fig:hybrid_combined}
\end{figure}

Visual comparisons between hybrid predictions and CFD ground truth are shown in Fig.~\ref{fig:hybrid_combined} on the X-Y plane at \( Z = 50\,\text{m} \), corresponding to the horizontal mid-plane of the domain. Each row displays, from left to right: CFD reference, hybrid model output, and absolute error.

\paragraph{3. Improvement Over Zeroth-Order Ansatz}

Compared to the zeroth-order ansatz, the hybrid model substantially reduces ringing artifacts and captures nonlinearities and fine-scale structures. This generalization holds across both test angles.

\begin{figure}[htbp]
  \centering

  \begin{subfigure}[b]{0.85\textwidth}
    \includegraphics[width=\textwidth]{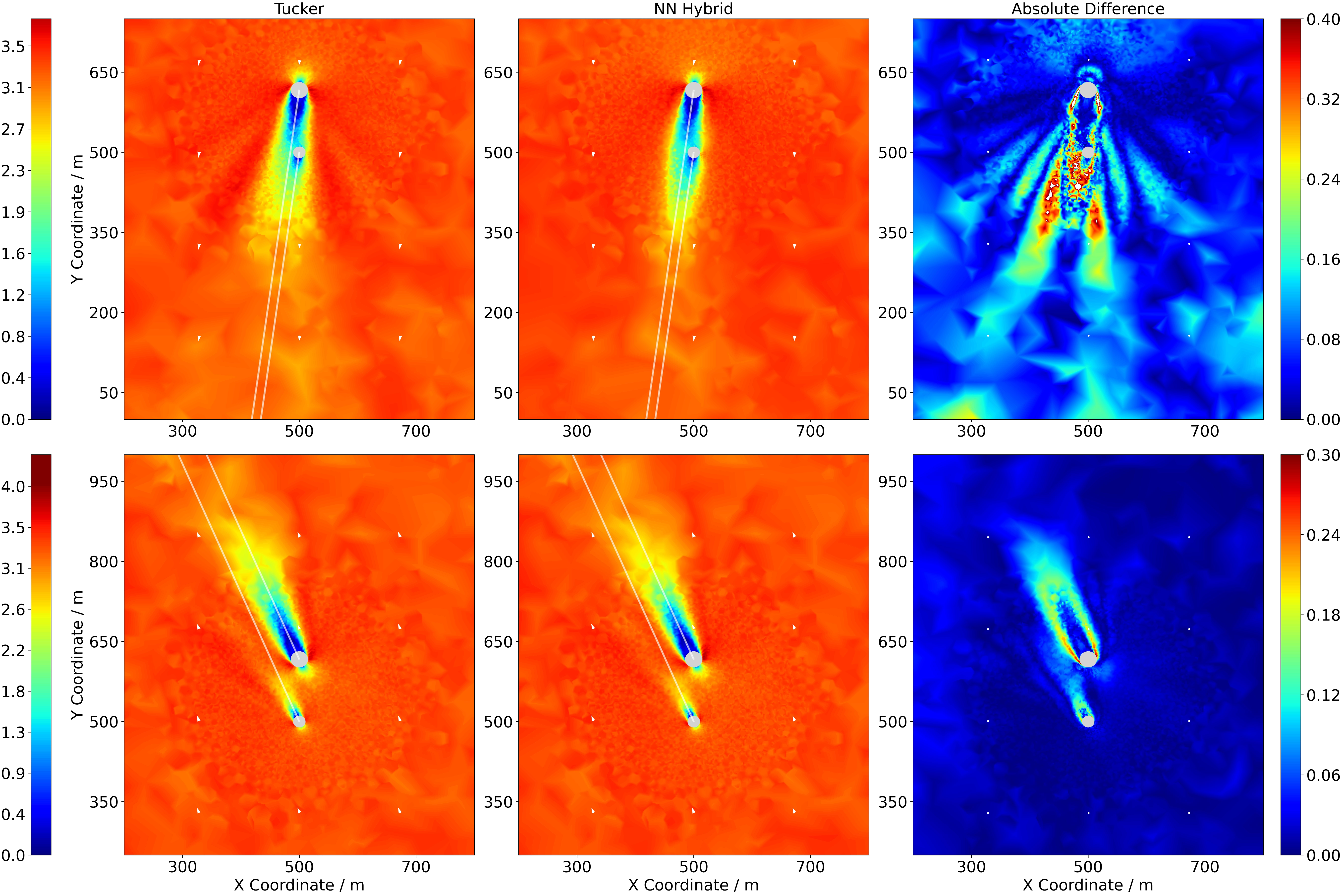}
    \caption{Velocity [$\mathrm{m}\,\mathrm{s}^{-1}$]}
    \label{fig:velocity_comp}
  \end{subfigure}
  \hfill
  \begin{subfigure}[b]{0.85\textwidth}
    \includegraphics[width=\textwidth]{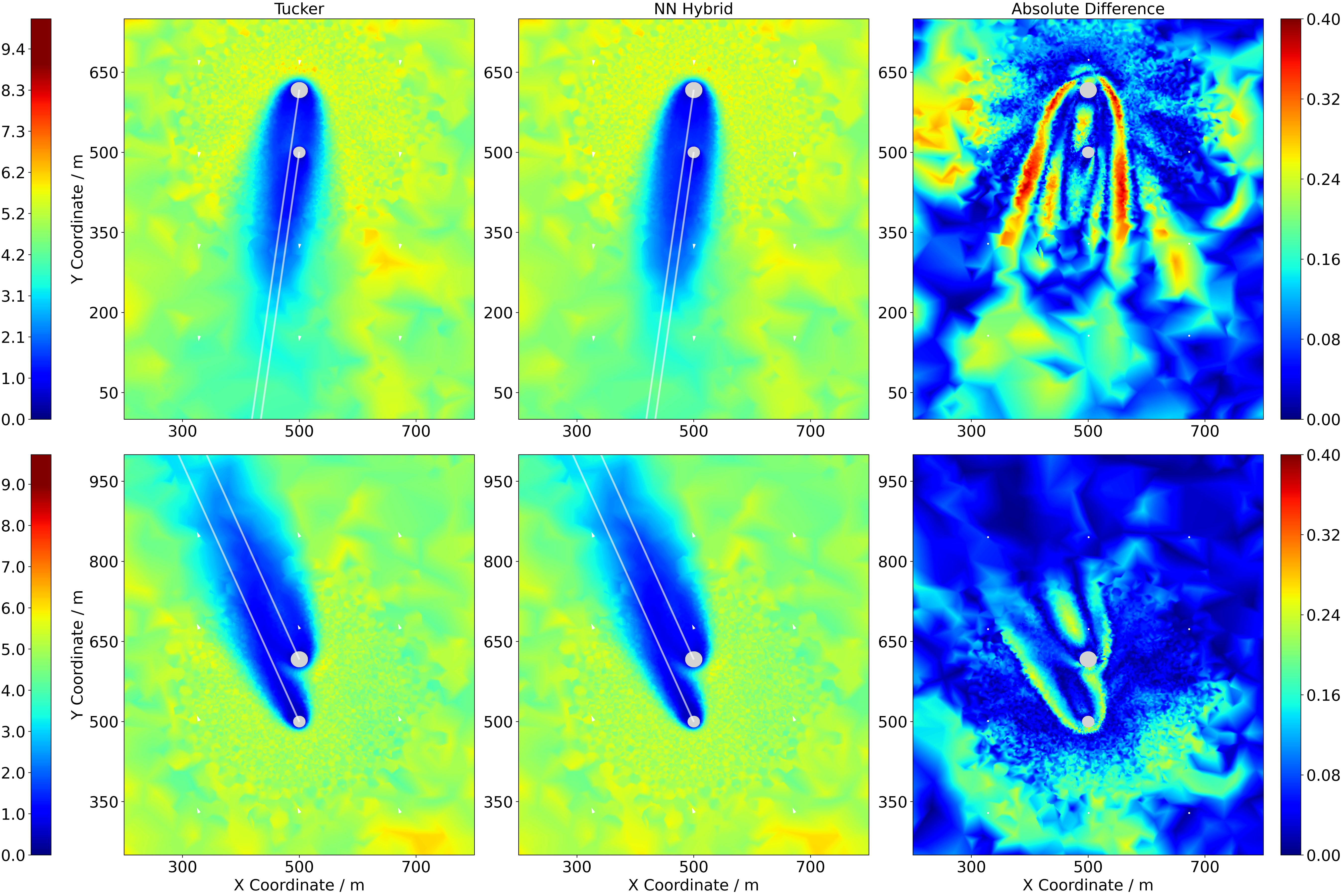}
    \caption{Eddy viscosity [$\mathrm{m}^2\,\mathrm{s}^{-1}$]}
    \label{fig:turbvisc_comp}
  \end{subfigure}

  \caption{Comparison of RANS field predictions between the Tucker-only model and the hybrid Tucker-NN model at $Z = 50\,\mathrm{m}$. Each subfigure shows the Tucker-only result (left), hybrid model prediction (middle), and absolute error (right) for velocity, eddy viscosity, and pressure fields.}
  \label{fig:comparison_combined}
\end{figure}

\paragraph{4. Training Convergence and Runtime}

\begin{figure}[htbp]
  \centering
  \includegraphics[width=1\textwidth]{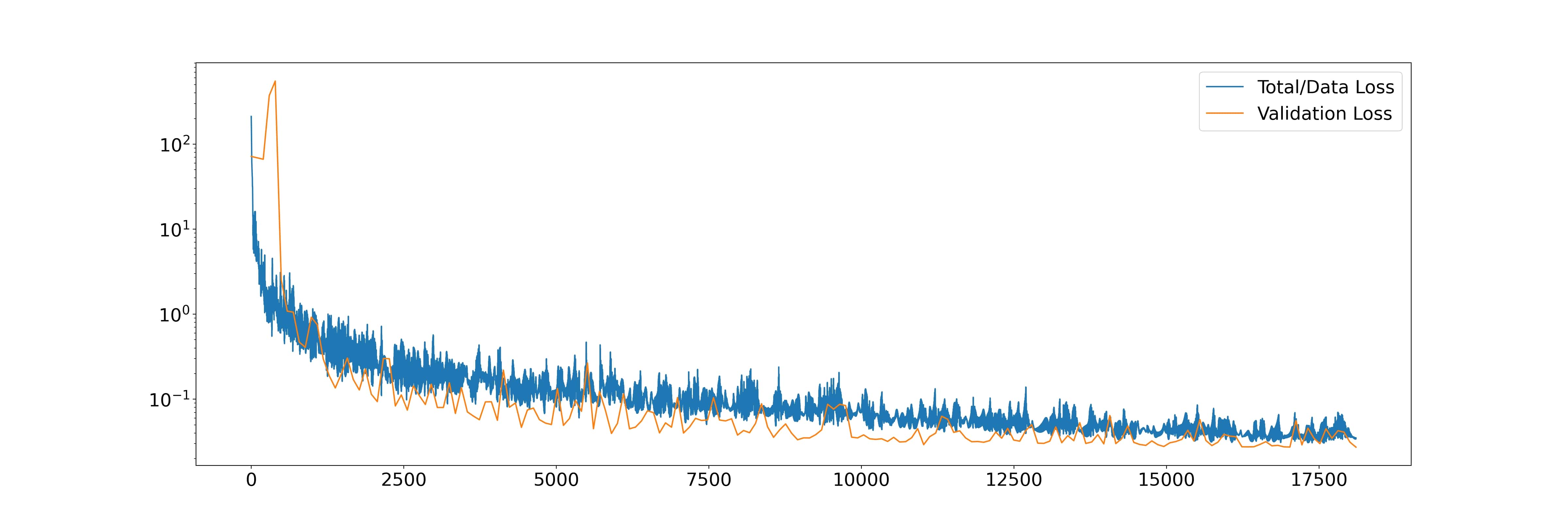}
  \caption{Training and validation loss over epochs for the hybrid Tucker-NN model. Stable convergence is observed.}
  \label{fig:hybrid_convergence}
\end{figure}

Fig.~\ref{fig:hybrid_convergence} shows stable convergence of training and validation losses. The model was trained on an RTX 3080 GPU, with each epoch taking approximately 0.45 seconds. Total training time was under 2 hours, including plot generation at every \checkinterval{}-epoch interval.

\paragraph{5. Summary}

The hybrid Tucker-NN model demonstrates strong predictive accuracy across all flow variables and angles. By incorporating a physics-informed ansatz and learning only the nonlinear residuals, the model achieves fast convergence and generalizes well. The architectural efficiency and reduced parameter count suggest potential for scalable deployment in urban wind simulations.

\section{Discussion and Conclusion}

This work evaluates two surrogate modeling approaches for urban wind prediction: a pure neural network model that establishes a reference benchmark for achievable accuracy, and a hybrid Tucker-NN architecture designed as a computationally efficient approximate alternative. We now interpret their quantitative and qualitative performance, theoretical implications, and computational trade-offs. The pure model remains fully feasible as a reference; we include it explicitly to define the accuracy benchmark against which the hybrid is compared.

\subsection{Comparative Accuracy: Metrics and Visual Insights}

Table~\ref{tab:diff_combined_mse_r2} summarizes relative $R^2$ differences of the hybrid model against the pure NN benchmark and the zeroth-order ansatz. The hybrid is near-benchmark relative to the pure NN and generally improves $u_y$ and $u_z$ over the ansatz, with small negatives in some components. All comparisons use the same 24-angle training dataset for both models.

\begin{table}[htbp]
\centering
\small
\caption{Percentage improvement of the Hybrid model over the Pure NN benchmark and the Zeroth-Order Ansatz at wind angles $7.5^\circ$ and $157.5^\circ$, for $R^2$. Positive values indicate the Hybrid performs better.}
\label{tab:diff_combined_mse_r2}
\begin{tabular}{lllcc}
\hline
\textbf{Angle} & \textbf{Metric} & \textbf{Variable} & \textbf{vs Pure NN (\%)} & \textbf{vs Ansatz (\%)} \\
\hline
\multirow{5}{*}{$7.5^\circ$}
 & \multirow{5}{*}{$R^2$}
  & $p$ & -0.21 & -0.21 \\
 & & $u_x$ & -2.03 & -1.42 \\
 & & $u_y$ & -0.37 & +0.13 \\
 & & $u_z$ & -1.79 & +0.64 \\
 & & $\nu_t$ & +0.06 & +0.06 \\
\hline
\multirow{5}{*}{$157.5^\circ$}
 & \multirow{5}{*}{$R^2$}
  & $p$ & -0.60 & -0.47 \\
 & & $u_x$ & -0.69 & +0.01 \\
 & & $u_y$ & -0.32 & +0.17 \\
 & & $u_z$ & -1.64 & +0.88 \\
 & & $\nu_t$ & -0.01 & -0.01 \\
\hline
\end{tabular}
\end{table}

\subsection{Discussion of Comparative Results}
The comparative metrics in Table~\ref{tab:diff_combined_mse_r2} do not exhibit a single monotonic trend across angles or variables. Relative to the pure NN benchmark, hybrid $R^2$ differences are small (within about 0--2\% across variables); relative to the zeroth-order ansatz, the hybrid generally improves $u_y$ and $u_z$ and is near parity for $u_x$ and $\nu_t$.

At $7.5^\circ$, the hybrid slightly trails the benchmark in $u_x$ and $u_z$ (about 1--2\%) while matching $\nu_t$ and marginally improving $p$ ($\sim$+0.2\%). Versus the ansatz, the hybrid yields improvements in $u_y$ and $u_z$ with a small deficit in $u_x$ and unchanged $\nu_t$. At $157.5^\circ$, the hybrid shows small negative deltas against the benchmark across $p$, $u_x$, $u_y$, and $u_z$ (within $\sim$0.4--1.8\%), and near-parity in $\nu_t$; versus the ansatz, it improves $u_y$ and $u_z$, is on par for $u_x$, and shows small negatives for $p$ and $\nu_t$.

These trends are consistent with the models’ design and the test geometry:
\begin{itemize}
  \item \textbf{Flow-regime heterogeneity}: The two test angles probe distinct regimes (shielding near $0^\circ$ vs exposure near $180^\circ$), which can favor different inductive biases across variables.
  \item \textbf{Metric sensitivity}: $R^2$ is sensitive to global offsets and variance capture. Pressure is affected by gauge shifts; eddy viscosity is localized and heavy-tailed, so small absolute errors can appear as modest $R^2$ deltas.
  \item \textbf{Ansatz proximity and residual learning}: Where the Tucker ansatz already explains most variance, residuals are small and relative gains appear modest; where Fourier interpolation introduces ringing near discontinuities, the hybrid’s nonlinearity yields localized improvements.
  \item \textbf{Representation effects}: The benchmark uses cell-centered fields while the hybrid uses nodal values; this pre-processing difference can slightly shift $R^2$ while preserving qualitative agreement and relative ranking.
  \item \textbf{Practical trade-off}: Given the $\sim$6$\times$ training speedup and reduced parameter count, the hybrid achieves near-benchmark accuracy for most variables and angles. The pure NN remains the appropriate choice when maximum fidelity is prioritized and compute budget permits; the hybrid is well-suited for time-critical or resource-constrained scenarios.
\end{itemize}

Taken together, the comparisons support three consistent conclusions:

\begin{enumerate}
    \item The hybrid model matches or approaches the benchmark accuracy established by the pure NN while dramatically reducing training cost, demonstrating that the accelerated approximate method successfully achieves reference-quality predictions.
    \item The zeroth-order ansatz achieves accuracy similar to the pure NN benchmark while the hybrid model corrects the ringing effects observed. This aligns with the hybrid's design: a physics-structured prior plus residual NN correction.
    \item Variability across angles and variables reflects inherent flow-regime differences and metric sensitivities rather than instability in the method.
\end{enumerate}

Finally, the visual assessments (Figures~\ref{fig:tucker_init}, \ref{fig:hybrid_combined}, and \ref{fig:comparison_combined}) corroborate the quantitative findings: the hybrid consistently suppresses angular ringing, preserves wake topology, and improves local fidelity in shear and recirculation zones, particularly at the more challenging $7.5^\circ$ case.

Visual comparisons (Fig.~\ref{fig:comparison_combined}) further highlight the hybrid model’s ability to suppress artifacts and recover physically plausible flow fields, even in challenging regions like wakes and shear layers.

\subsection{Physics Consistency and Flow Structures}

Although neither model explicitly enforces conservation laws, both are trained on data that satisfy mass and momentum conservation. As a result, the learned velocity fields retain physically consistent behavior. Interestingly, both models exhibit smoother velocity divergence profiles than the CFD ground truth, suggesting a beneficial denoising effect.

The zeroth-order ansatz itself satisfies physics constraints by construction. Its inclusion as a prior in the hybrid model provides an inductive bias that helps preserve structure and flow symmetry, especially in wake interactions and eddy distributions.

\subsection{Computational Cost and Scalability}

The hybrid architecture offers a favorable trade-off between fidelity and efficiency:
\begin{itemize}
  \item \textbf{Pure NN (benchmark reference):} High accuracy establishing the target performance level, but computationally demanding training ($\sim$2.82 sec/epoch on RTX 3080 GPU).
  \item \textbf{Tucker-only:} Near-instant evaluation but poor local fidelity.
  \item \textbf{Hybrid (accelerated approximate):} Achieves benchmark-competitive accuracy with fast training ($\sim$0.4 sec/epoch on an RTX 3080 GPU), totaling  $\sim$2 hours for complete training and visualization.
\end{itemize}

The hybrid model achieves significant training speedup compared to the pure NN benchmark. In our experiments on an RTX 3080 GPU, per-epoch times were approximately 2.82 s for the pure NN and 0.4 s for the hybrid. While absolute times vary by hardware and implementation details, the gains stem from architectural efficiency: reduced parameter count (16,197 vs. 50,949 parameters) and data compression through Tucker decomposition (100,680 nodal points vs. 352,071 cell-centered values in the pure NN). The 3.1-fold parameter reduction and 3.5-fold data reduction directly translate to proportionally lower memory requirements for model weights, gradients, and training batches, which is hardware-independent and suggests strong potential scalability to larger computational domains.

For real-time applications, inference time is critical. Once trained, both the pure NN and hybrid models perform inference (forward pass) in milliseconds per query on GPU, orders of magnitude faster than CFD solves which require hours to days. The hybrid model's inference involves: (1) Tucker reconstruction via tensor contraction, and (2) a single forward pass through the residual network. Both operations are computationally inexpensive compared to the training phase, making the approach suitable for operational scenarios requiring rapid wind field updates in response to changing meteorological conditions.

\subsection{Limitations and Future Work}

Despite its strengths, the proposed method has two key limitations:
\begin{enumerate}
  \item \textbf{Angular Sampling:} Ansatz currently assumes periodic angular coverage with constant $\Delta \theta$. Future work could explore cases where $\Delta \theta$ varies and does not span the full $2\pi$ region.
  \item \textbf{Parameter Dimensionality:} Current variation is restricted to inlet angle. Extending to include amplitude, inlet wind angle variation with respect to $z$, and time-averaged velocity fluctuations will require higher-order tensor decompositions.
  \item \textbf{Fourier Dependence:} Method is fully reliant on Fourier interpolation; introducing a non-periodic parameter would break the formulation.
\end{enumerate}

To address these challenges, future work will explore interpolation methods beyond Fourier, such as polynomial or spline-based approaches, together with adaptive sampling strategies to improve generalization at manageable cost.

\subsection{Conclusion}

We have presented a hybrid surrogate modeling framework that combines the interpretability and speed of tensor decomposition with the flexibility of neural residual learning. By establishing a pure neural network model as an accuracy benchmark (Section~\ref{sec:purenn}), we demonstrated that the hybrid architecture (Section~\ref{sec:hybridnn}), an accelerated approximate alternative, successfully achieves reference-quality predictions while offering substantial computational advantages:
\begin{itemize}
  \item Matches or approaches benchmark accuracy established by the pure NN model
  \item Suppresses spurious flow artifacts and preserves wake dynamics
  \item Achieves sub-second per-epoch training times ($\sim$6$\times$ faster than the benchmark)
  \item Reduces parameter count and memory footprint, suggesting scalability to larger domains
\end{itemize}

The hybrid Tucker-NN model bridges data-driven regression with physically structured modeling, demonstrating that structured decomposition can serve as an effective accelerator for neural network-based wind field interpolation without sacrificing accuracy. The combination of benchmark-competitive performance and dramatic computational efficiency suggests strong potential for operational urban wind simulation applications. Future work will extend the method to more complex geometries, additional environmental variables, and broader classes of PDE systems.

\section{Data and Code Availability}
The CFD datasets and Python code implementing the hybrid Tucker-NN model will be made available upon reasonable request to the corresponding author. The CFD simulations were performed using the open-source Code\_Saturne software (\url{https://www.code-saturne.org/}). The neural network implementations utilize PyTorch, with tensor decompositions performed using TensorLy. All software dependencies (NumPy, SciPy, PyTorch, TensorLy) are publicly available open-source libraries.

\section{Acknowledgment}
This research was initiated by and partially carried within the Descartes programme supported by the National Research Foundation, Prime Minister's Office, Singapore under its Campus for Research Excellence and Technological Enterprise (CREATE). One of the authors (AS) is grateful to CNRS@CREATE for full support. One of the authors (AS) is grateful to Professor Xavier Garbet (NTU) for full support from Dec 2024 to present. One of the authors (CG) is grateful to CNRS@CREATE for partial support. All of the authors acknowledge Dr. Wang Zhe's contributions in the early stages of this project.

\bibliographystyle{unsrt}
\bibliography{references}

\begin{thebibliography}{10}

\bibitem{britter2003flow}
R.~E. Britter and S.~R. Hanna.
\newblock Flow and dispersion in urban areas.
\newblock {\em Annual Review of Fluid Mechanics}, 35:469--496, 2003.

\bibitem{franke2007best}
J.~Franke, A.~Hellsten, H.~Schlünzen, and B.~Carissimo.
\newblock Best practice guideline for the cfd simulation of flows in the urban
  environment (cost 732).
\newblock Technical report, 2007.

\bibitem{batchelor1967}
G.~K. Batchelor.
\newblock {\em An Introduction to Fluid Dynamics}.
\newblock Cambridge University Press, 1967.

\bibitem{temam2001}
R.~Temam.
\newblock {\em Navier-Stokes Equations: Theory and Numerical Analysis}.
\newblock American Mathematical Society, 2001.

\bibitem{moin1998}
P.~Moin and K.~Mahesh.
\newblock Direct numerical simulation: a tool in turbulence research.
\newblock {\em Annual Review of Fluid Mechanics}, 30(1):539--578, 1998.

\bibitem{Pope2000}
Stephen~B. Pope.
\newblock {\em Turbulent Flows}.
\newblock Cambridge University Press, Cambridge, UK, 2000.

\bibitem{reynolds1895}
O.~Reynolds.
\newblock On the dynamical theory of incompressible viscous fluids and the
  determination of the criterion.
\newblock {\em Philosophical Transactions of the Royal Society of London A},
  186:123--164, 1895.

\bibitem{berkooz1993proper}
Gal Berkooz, Philip Holmes, and John~L Lumley.
\newblock The proper orthogonal decomposition in the analysis of turbulent
  flows.
\newblock {\em Annual Review of Fluid Mechanics}, 25(1):539--575, 1993.

\bibitem{du2013pod}
J.~Du, F.~Fang, C.~C. Pain, I.~M. Navon, J.~Zhu, and D.~A. Ham.
\newblock Pod reduced-order unstructured mesh modeling applied to 2d and 3d
  fluid flow.
\newblock {\em Computers and Mathematics with Applications}, 65:362--369, 2013.

\bibitem{amsallem2008interpolation}
David Amsallem and Charbel Farhat.
\newblock Interpolation method for adapting reduced-order models and
  application to aeroelasticity.
\newblock {\em AIAA Journal}, 46(7):1803--1813, 2008.

\bibitem{raissi2019physics}
Maziar Raissi, Paris Perdikaris, and George~Em Karniadakis.
\newblock Physics-informed neural networks: A deep learning framework for
  solving forward and inverse problems involving nonlinear partial differential
  equations.
\newblock {\em Journal of Computational Physics}, 378:686--707, 2019.

\bibitem{karniadakis2021}
G.~E. Karniadakis, I.~G. Kevrekidis, L.~Lu, P.~Perdikaris, S.~Wang, and
  L.~Yang.
\newblock Physics-informed machine learning.
\newblock {\em Nature Reviews Physics}, 3(6):422--440, 2021.

\bibitem{cuomo2022}
S.~Cuomo, V.~S. Di~Cola, F.~Giampaolo, G.~Rozza, M.~Raissi, and F.~Piccialli.
\newblock Scientific machine learning through physics--informed neural
  networks: Where we are and what's next.
\newblock {\em Journal of Scientific Computing}, 92(3):88, 2022.

\bibitem{lee2020data}
Sangseung Lee and Donghyun You.
\newblock Data-driven prediction of unsteady flow over a circular cylinder
  using deep learning.
\newblock {\em Journal of Fluid Mechanics}, 879:217--254, 2019.

\bibitem{Kolda2009}
Tamara~G. Kolda and Brett~W. Bader.
\newblock Tensor decompositions and applications.
\newblock {\em SIAM Review}, 51(3):455--500, 2009.

\bibitem{hitchcock1927}
F.~L. Hitchcock.
\newblock The expression of a tensor or a polyadic as a sum of products.
\newblock {\em Journal of Mathematics and Physics}, 6(1--4):164--189, 1927.

\bibitem{carroll1970}
J.~D. Carroll and J.~J. Chang.
\newblock Analysis of individual differences in multidimensional scaling via an
  n-way generalization of "eckart-young" decomposition.
\newblock {\em Psychometrika}, 35(3):283--319, 1970.

\bibitem{reiss2018shifted}
Julius Reiss, Philipp Schulze, J{\"o}rg Sesterhenn, and Volker Mehrmann.
\newblock The shifted proper orthogonal decomposition: A mode decomposition for
  multiple transport phenomena.
\newblock {\em SIAM Journal on Scientific Computing}, 40(3):A1322--A1344, 2018.

\bibitem{peherstorfer2018survey}
Benjamin Peherstorfer, Karen Willcox, and Max Gunzburger.
\newblock Survey of multifidelity methods in uncertainty propagation,
  inference, and optimization.
\newblock {\em SIAM Review}, 60(3):550--591, 2018.

\bibitem{kutz2016dmd}
J.~Nathan Kutz, Steven~L. Brunton, Bingni~W. Brunton, and Joshua~L. Proctor.
\newblock {\em Dynamic Mode Decomposition: Data-Driven Modeling of Complex
  Systems}.
\newblock SIAM, 2016.

\bibitem{novikov2015tensorizing}
Alexander Novikov, Dmitry Podoprikhin, Anton Osokin, and Dmitry~P. Vetrov.
\newblock Tensorizing neural networks.
\newblock In {\em Advances in Neural Information Processing Systems},
  volume~28, pages 442--450, 2015.

\bibitem{kossaifi2019tensor}
Jean Kossaifi, Yannis Panagakis, Anima Anandkumar, and Maja Pantic.
\newblock Tensor regression networks.
\newblock {\em Neural Information Processing Systems (NeurIPS)}, pages 1--11,
  2019.

\bibitem{jones1972}
W.~P. Jones and B.~E. Launder.
\newblock The prediction of laminarization with a two-equation model of
  turbulence.
\newblock {\em International Journal of Heat and Mass Transfer},
  15(2):301--314, 1972.

\bibitem{launder1974numerical}
B.~E. Launder and D.~B. Spalding.
\newblock The numerical computation of turbulent flows.
\newblock {\em Computer Methods in Applied Mechanics and Engineering},
  3:269--289, 1974.

\bibitem{rodi1993}
W.~Rodi.
\newblock {\em Turbulence Models and Their Application in Hydraulics}.
\newblock CRC Press, 1993.

\bibitem{archambeau2004codesaturne}
Fran\c{c}ois Archambeau, Najib M\'echitoua, and Marcel Sakiz.
\newblock Code\_saturne: A finite-volume code for the computation of turbulent
  incompressible flows.
\newblock {\em International Journal on Finite Volumes}, 1(1), 2004.

\bibitem{ferziger2002}
J.~H. Ferziger and M.~Perić.
\newblock {\em Computational Methods for Fluid Dynamics}.
\newblock Springer, 2002.

\bibitem{salome}
Salome platform: Open-source integration platform for numerical simulation.
\newblock \url{https://www.salome-platform.org}.
\newblock Accessed 2025-09-12.

\bibitem{lacome2017guide}
Jean-Marc Lacome and Benjamin Truchot.
\newblock Guide de bonnes pratiques pour la réalisation de modélisations {3D}
  pour des scénarios de dispersion atmosphérique en situation accidentelle.
\newblock Technical Report DRA-15-148997-06852A, INERIS, 2017.

\bibitem{panofsky1984}
H.~A. Panofsky and J.~A. Dutton.
\newblock {\em Atmospheric Turbulence: Models and Methods for Engineering
  Applications}.
\newblock John Wiley \& Sons, 1984.

\bibitem{stull1988}
R.~B. Stull.
\newblock {\em An Introduction to Boundary Layer Meteorology}.
\newblock Springer, 1988.

\bibitem{achenbach1974vortex}
Elmar Achenbach.
\newblock Vortex shedding from spheres.
\newblock {\em Journal of Fluid Mechanics}, 62(2):209--221, January 1974.

\bibitem{leishman2023bluff}
J.~Gordon Leishman.
\newblock Bluff body flows.
\newblock {\em Introduction to Aerospace Flight Vehicles}, 2023.

\bibitem{roshko1961experiments}
Anatol Roshko.
\newblock Experiments on the flow past a circular cylinder at very high
  reynolds number.
\newblock {\em Journal of Fluid Mechanics}, 10(3):345--356, 1961.

\bibitem{williamson1996}
C.~H.~K. Williamson.
\newblock Vortex dynamics in the cylinder wake.
\newblock {\em Annual Review of Fluid Mechanics}, 28(1):477--539, 1996.

\bibitem{Ahrens2005}
James Ahrens, Berk Geveci, and Charles Law.
\newblock Paraview: An end-user tool for large data visualization.
\newblock In Charles~D. Hansen and Chris~R. Johnson, editors, {\em The
  Visualization Handbook}, pages 717--731. Elsevier, 2005.

\bibitem{Clevert2016}
Djork-Arné Clevert, Thomas Unterthiner, and Sepp Hochreiter.
\newblock Fast and accurate deep network learning by exponential linear units
  (elus).
\newblock In {\em Proceedings of the International Conference on Learning
  Representations (ICLR)}, 2016.

\bibitem{Rumelhart1986}
David~E. Rumelhart, Geoffrey~E. Hinton, and Ronald~J. Williams.
\newblock Learning representations by back‐propagating errors.
\newblock In {\em Proceedings of the Royal Society B}, volume 323, pages
  533--544, 1986.

\bibitem{lecun2012}
Y.~A. LeCun, L.~Bottou, G.~B. Orr, and K.~R. Müller.
\newblock Efficient backprop.
\newblock In {\em Neural Networks: Tricks of the Trade}, pages 9--48. 2012.

\bibitem{KingmaBa2015}
Diederik~P. Kingma and Jimmy Ba.
\newblock Adam: A method for stochastic optimization.
\newblock In {\em Proceedings of the International Conference on Learning
  Representations (ICLR)}, 2015.

\bibitem{eckart1936}
C.~Eckart and G.~Young.
\newblock The approximation of one matrix by another of lower rank.
\newblock {\em Psychometrika}, 1(3):211--218, 1936.

\bibitem{DeLathauwer2000}
Lieven De~Lathauwer, Bart De~Moor, and Joos Vandewalle.
\newblock A multilinear singular value decomposition.
\newblock {\em SIAM Journal on Matrix Analysis and Applications},
  21(4):1253--1278, 2000.

\bibitem{Kossaifi2019}
Jean Kossaifi, Yannis Panagakis, Anima Anandkumar, and Maja Pantic.
\newblock {TensorLy}: Tensor learning in python.
\newblock {\em Journal of Machine Learning Research}, 20(26):1--6, 2019.

\bibitem{gibbs1899fourier}
J.~Willard Gibbs.
\newblock Fourier's series.
\newblock {\em Nature}, 59(1522):606, 1899.

\bibitem{gottlieb1997}
D.~Gottlieb and C.~W. Shu.
\newblock On the gibbs phenomenon and its resolution.
\newblock {\em SIAM Review}, 39(4):644--668, 1997.

\bibitem{hewitt1979}
E.~Hewitt and R.~E. Hewitt.
\newblock The gibbs-wilbraham phenomenon: An episode in fourier analysis.
\newblock {\em Archive for History of Exact Sciences}, 21(2):129--160, 1979.

\bibitem{fix1951}
E.~Fix and J.~L. Hodges~Jr.
\newblock Discriminatory analysis-nonparametric discrimination: consistency
  properties.
\newblock Technical report, USAF School of Aviation Medicine, 1951.

\bibitem{cover1967}
T.~Cover and P.~Hart.
\newblock Nearest neighbor pattern classification.
\newblock {\em IEEE Transactions on Information Theory}, 13(1):21--27, 1967.

\bibitem{Buhmann2003}
B.~M. Buhmann.
\newblock {\em Radial Basis Functions: Theory and Implementations}.
\newblock Cambridge University Press, 2003.

\bibitem{Wendland2004}
H.~Wendland.
\newblock {\em Scattered Data Approximation}.
\newblock Cambridge University Press, 2004.

\bibitem{Fasshauer2007}
G.~E. Fasshauer.
\newblock {\em Meshfree Approximation Methods with MATLAB}.
\newblock World Scientific, 2007.

\bibitem{Glorot2011}
Xavier Glorot, Antoine Bordes, and Yoshua Bengio.
\newblock Deep sparse rectifier neural networks.
\newblock In {\em Proceedings of the Fourteenth International Conference on
  Artificial Intelligence and Statistics}, pages 315--323. JMLR Workshop and
  Conference Proceedings, 2011.

\bibitem{nair2010}
V.~Nair and G.~E. Hinton.
\newblock Rectified linear units improve restricted boltzmann machines.
\newblock In {\em Proceedings of the 27th International Conference on Machine
  Learning}, pages 807--814, 2010.

\bibitem{picard1893}
E.~Picard.
\newblock Sur l'application des méthodes d'approximations successives à
  l'étude de certaines équations différentielles ordinaires.
\newblock {\em Journal de Mathématiques Pures et Appliquées}, 9:217--271,
  1893.

\bibitem{arnold1992ode}
Vladimir~I. Arnold.
\newblock {\em Ordinary Differential Equations}.
\newblock Springer, Berlin, 1992.

\bibitem{banach1922}
Stefan Banach.
\newblock Sur les opérations dans les ensembles abstraits et leur application
  aux équations intégrales.
\newblock {\em Fundamenta Mathematicae}, 3:133--181, 1922.

\bibitem{granas2003fixed}
Andrzej Granas and James Dugundji.
\newblock {\em Fixed Point Theory}.
\newblock Springer, New York, 2003.

\bibitem{kumari2021r2}
Sushma Kumari and Deepak Dogra.
\newblock A differentiable r-squared loss for regression, 2021.

\bibitem{Saxe2014}
Andrew~M Saxe, James~L McClelland, and Surya Ganguli.
\newblock Exact solutions to the nonlinear dynamics of learning in deep linear
  neural networks.
\newblock {\em International Conference on Learning Representations (ICLR)},
  2014.

\bibitem{fejer1900}
Lipót Fejér.
\newblock Über die fourierschen reihen.
\newblock {\em Mathematische Annalen}, 58(3):501--569, 1904.

\bibitem{cesaro1888}
Ernesto Cesàro.
\newblock Sur la multiplication des séries.
\newblock {\em Bulletin de la Société Mathématique de France}, 16:304--312,
  1888.

\end{thebibliography}

\appendix
\section{Derivation of Eq.~\eqref{eq:fejernn}}\label{sec:fejer}

Below we give a self-contained derivation showing that the \emph{Cesàro mean} of the modified Dirichlet kernels
\[
K_j(\theta) = \frac{1}{2}\left(1+ \frac{\sin\left(M (\theta-\theta_j) - \frac{\theta-\theta_j}{2}\right)}{\sin\left(\frac{\theta-\theta_j}{2}\right)}\right)
\]

yields a Fejér-like kernel. Setting $\phi = \theta - \theta_j$ and $x = \tfrac{\phi}{2}$ for clarity:
\[
K_M(\phi) = \frac{1}{2}\left(1+ \frac{\sin\left((2M-1)x\right)}{\sin x}\right)
\]

\medskip

\begin{proof}

\begin{enumerate}
 \item \textbf{Cesàro mean of the kernels}\\
 By definition, the Cesàro mean is:
 \[
  \tilde{K}_N(\phi) := \frac{1}{N}\sum_{M=1}^{N}K_M(\phi) = \frac{1}{2} + \frac{1}{2N \sin x} \sum_{M=1}^N \sin[ (2M -1) x]
 \]

 \item \textbf{Compute the sum} \\
    Rewriting the summand as a sum of complex exponentials and performing the finite geometric sum,
     \[
      \sum_{M=1}^N \sin[ (2M -1) x] = \tfrac{i}{2}\sum_{M=1}^N \left[e^{-ix(2M-1))} - e^{+ix(2M-1)}\right] = \frac{\sin^2(Nx)}{\sin x}
     \]
     where the first and last step involves the identity $\sin x = \tfrac{1}{2}(e^{-ix} - e^{ix})$
     
 \item \textbf{Final Result} \\
 Substituting the result, 
 \[
  \tilde{K}_N(\phi) = \frac{1}{2} \left\{1 + \frac{1}{N} \frac{\sin^2(Nx)}{\sin^2(x)} \right\}
 \]

 where we recover the $\frac{\sin^2(.)}{\sin^2(.)}$  structure reminiscent of the Fejér kernel \cite{fejer1900}. The method uses the definition of a Cesàro summation \cite{cesaro1888} which averages frequency contributions to reduce oscillations.
\end{enumerate}
\end{proof}

\bigskip
\subsection{Definition of the Cesàro Mean}\label{def:cesaro}
Given a sequence of partial sums \(\{S_n\}_{n\ge0}\), the \emph{Cesàro mean} of order 0 (or \((C,1)\) mean) is
\[
 \sigma_N
 =\frac{1}{N+1}\sum_{n=0}^{N}S_n.
\]

\subsection{Scaling to Match the Dirichlet Peak}\label{sec:scaling}

In the context of this paper, one wishes the product of the Dirichlet kernel and the Fejér-like kernel to maintain the same maximum value as $K_M(\phi)$ alone.

First, evaluate each kernel at the limit where $\phi \to 0$:
\begin{equation}
K_M(0) = \frac{1}{2}\left(1 + \lim_{\phi \to 0}\frac{\sin\left((2M-1)\frac{\phi}{2}\right)}{\sin\left(\frac{\phi}{2}\right)}\right) = \frac{1}{2}(1 + (2M-1)) = M
\end{equation}

For the Fejér-like kernel, the dominant term at $\phi = 0$ gives:
\begin{equation}
\lim_{\phi \to 0} \tilde{K}_N(\phi) = \frac{1}{2} \cdot \left\{1 + \tfrac{1}{N} (N^2) \right\} = \frac{N+1}{2}
\end{equation}

Thus the unscaled product peaks at:
\begin{equation}
P_{M,N}(0) = K_M(0) \cdot \tilde{K}_N(0) = \frac{M(N+1)}{2}
\end{equation}

\textbf{Correct scaling:} To preserve the peak height of $M$, we scale the product by $\frac{2}{N+1}$:

\begin{equation}
\boxed{\widehat{P}_{M,N}(\phi) = \frac{2}{N+1} \cdot K_M(\phi) \cdot \tilde{K}_N(\phi)}
\end{equation}

This gives:
\begin{equation}
\widehat{P}_{M,N}(0) = \frac{2}{N+1} \cdot M \cdot \frac{N+1}{2} = M
\end{equation}

\end{document}